\documentclass[sigconf]{acmart}

\usepackage{xpunctuate} %
\usepackage{enumitem} %
\usepackage{tikz} %
\setlist{nosep}
\setenumerate{nosep}
\usepackage{array}
\usepackage{booktabs}
\usepackage{makecell}

\AtBeginDocument{%
  }

\setcopyright{acmlicensed}
\copyrightyear{2018}
\acmYear{2018}
\acmDOI{XXXXXXX.XXXXXXX}
\acmConference[Conference acronym 'XX]{Make sure to enter the correct
  conference title from your rights confirmation email}{June 03--05,
  2018}{Woodstock, NY}
\acmISBN{978-1-4503-XXXX-X/2018/06}

\begin{document}

\title[VideoTrail: Interactive Video Foraging Through Hierarchical Visualization]{VideoTrail: Interactive Video Foraging Through \\ Hierarchical Visualization}

\author{Catherine Yeh}
\orcid{0009-0007-0429-4770}
\affiliation{%
  \institution{Harvard University}
  \city{Boston}
  \state{MA}
  \country{USA}}
\email{catherineyeh@g.harvard.edu}

\author{Bryan Wang}
\orcid{0000-0001-9016-038X}
\affiliation{%
  \institution{Adobe Research}
  \city{Seattle}
  \state{WA}
  \country{USA}}
\email{bryanw@adobe.com}

\author{Fernanda Vi\'egas}
\authornote{FV and MW are also with Google DeepMind, but this work was done at Harvard.}
\orcid{0000-0002-4951-4008}
\affiliation{%
  \institution{Harvard University}
  \city{Boston}
  \state{MA}
  \country{USA}}
\email{fernanda@g.harvard.edu}

\author{Martin Wattenberg}
\authornotemark[1]
\orcid{0000-0003-0904-4862}
\affiliation{%
  \institution{Harvard University}
  \city{Boston}
  \state{MA}
  \country{USA}}
\email{wattenberg@g.harvard.edu}

\renewcommand{\shortauthors}{Yeh et al.}

\newcommand{\ie}{{i.e.,}\xspace}
\newcommand{\eg}{{e.g.,}\xspace}
\newcommand{\ea}{{et~al\xperiod}\xspace}
\newcommand{\aka}{{a.k.a.}\xspace}
\newcommand{\etc}{{etc\xperiod}\xspace}
\newcommand{\etal}{{et al\xperiod}\xspace}
\newcommand{\vs}{{vs.}\xspace}

\newcommand{\todo}[1]{}
\newcommand{\cy}[1]{}

\newcommand{\edit}[1]{{\textcolor{black}{#1}\normalfont}}

\newcommand{\system}{\textsc{VideoTrail}\xspace} %

\newcommand{\location}{Harvard University\xspace}

\renewcommand{\sectionautorefname}{Sec.}
\renewcommand{\subsectionautorefname}{Sec.}
\renewcommand{\subsubsectionautorefname}{Sec.}
\renewcommand{\appendixautorefname}{App.}
\renewcommand{\figureautorefname}{Fig.}
\renewcommand{\tableautorefname}{Tab.}

\newcommand{\myquote}[1]{\emph{``#1''}}
\newcommand{\boldni}[1]{\noindent{\textbf{#1}}}
\newcommand{\boldpar}[1]{\paragraph{\textbf{#1}}}

\newcommand{\circled}[1]{%
    \tikz[baseline=(char.base)]{
        \node[shape=circle, fill=black, inner sep=1.5pt, text=white] (char) {\small\sffamily\bfseries#1};
    }%
}

\newcommand{\qa}[2]{\textbf{#1}\par #2\vspace{0.5em}\par}

\newcommand{\Contextualize}{\textbf{D1: \textit{Contextualize}}}
\newcommand{\Adapt}{\textbf{D2: \textit{Adapt}}}
\newcommand{\Balance}{\textbf{D3: \textit{Balance}}}
\newcommand{\Control}{\textbf{D4: \textit{Control}}}

\definecolor{PurpleColor}{HTML}{74628E}
\definecolor{DarkPurpleColor}{HTML}{2F0983}
\definecolor{OrangeColor}{HTML}{F0A04C}
\definecolor{DarkOrangeColor}{HTML}{D34600}
\definecolor{BlueColor}{HTML}{8B9CBD}
\definecolor{DialogueColor}{HTML}{01441B}
\definecolor{VisualColor}{HTML}{8B74B6}

\definecolor{visual}{HTML}{47645A}
\definecolor{dialogue}{HTML}{244F7E}
\definecolor{sound}{HTML}{71336D}
\definecolor{mood}{HTML}{8B5A26}
\definecolor{theme}{HTML}{9A7413}

\newcommand{\sig}[1]{\textbf{#1}}          %
\newcommand{\ns}[1]{#1}                    %
\newcommand{\cell}[5]{%
  B: $M\!=\!#1$\newline
  VT: $M\!=\!#2$\newline
  $\Delta\!=\!#3$\newline
  $W\!=\!#4$\newline
  #5%
}
\renewcommand\theadfont{\bfseries}
\renewcommand\theadalign{l}
\renewcommand\cellalign{tl}

\newcommand{\figureTeaser}{
\begin{teaserfigure}
  \includegraphics[width=\linewidth]{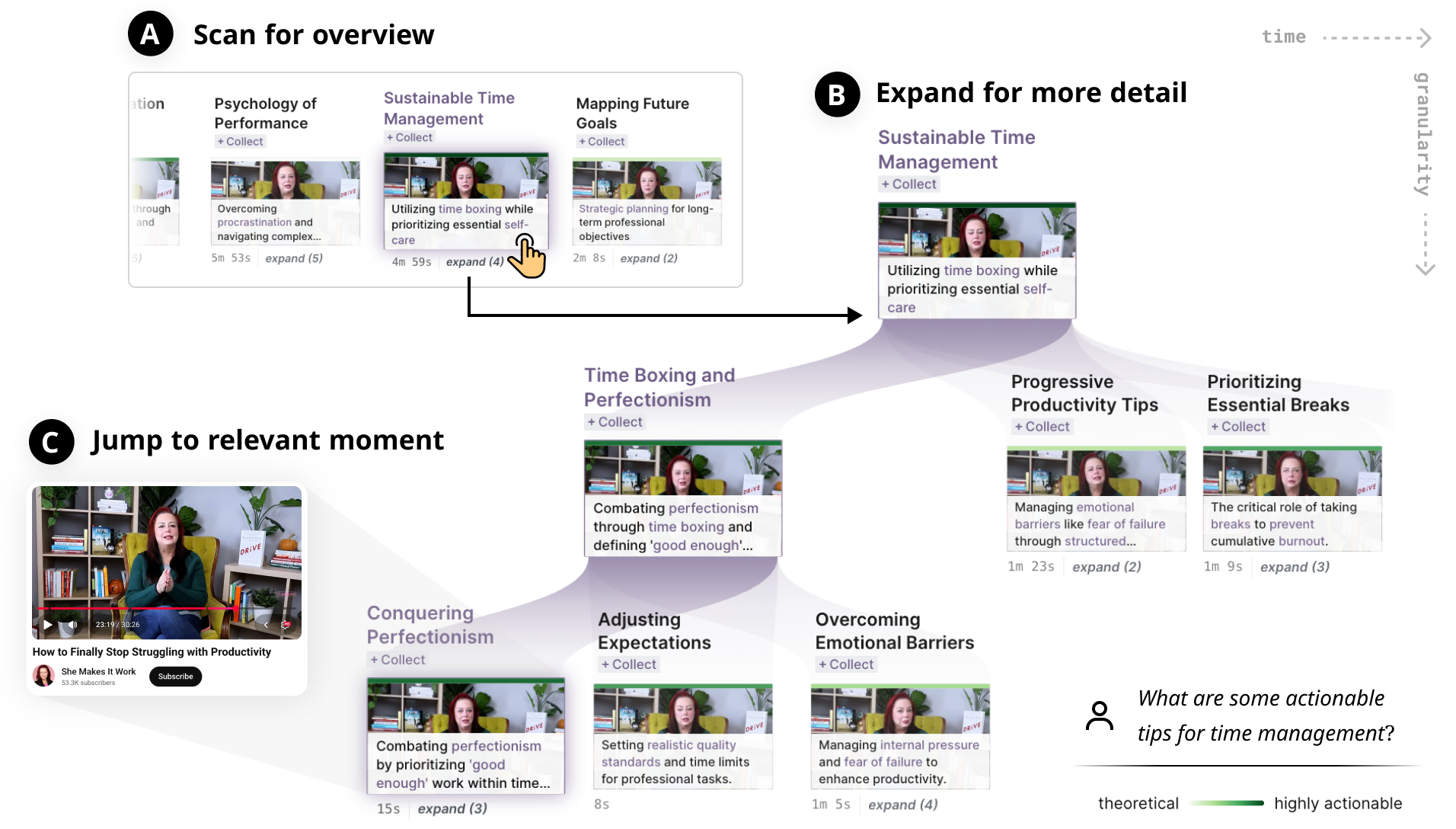}
  \caption{\textbf{\system generates interactive, hierarchical visualizations of YouTube videos to help users forage for information.} Here, a viewer seeking time-management tips \textbf{(A)} scans a high-level overview of segments; each includes a title, keyframe, summary, and colorbar encoding a video-specific semantic dimension (\ie actionability). \textbf{(B)} Expanding a segment reveals finer-grained subsegments and additional layers of the hierarchy. \textbf{(C)} Selecting a segment jumps to the corresponding moment in the video. Video source: \href{https://www.youtube.com/watch?v=W3RLUa4nJas}{How to Finally Stop Struggling with Productivity} by \textit{She Makes It Work} (Creative Commons license).}
  \Description{Three-panel diagram of VideoTrail's hierarchical video map. Panel A shows a chronological overview of video segment cards; panel B shows a segment expanding into finer-grained subsegments across three layers; and panel C shows the video player at the selected moment. Time progresses from left to right, granularity increases from top to bottom, and a scale encodes actionability from theoretical to highly actionable.}
  \label{fig:teaser}
\end{teaserfigure}
}

\newcommand{\figurePipeline}{
\begin{figure*}[t]
  \includegraphics[width=\textwidth]{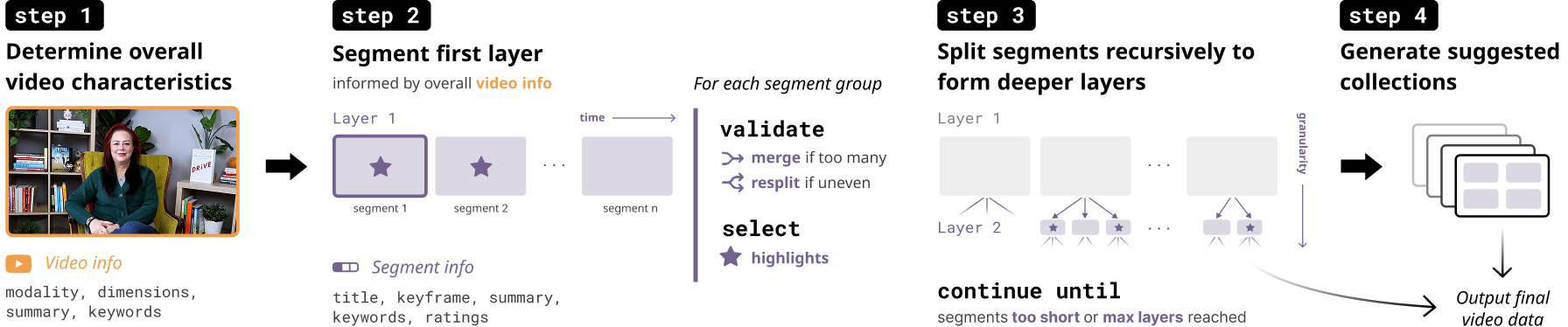}
  \caption{\textbf{Overview of our AI-powered data pipeline.} We collect overall characteristics about the input video, followed by recursive segmentation to form the hierarchical data structure. The pipeline also returns a list of suggested segment collections.}
  \Description{Pipeline for extracting hierarchical video structure. The diagram shows four steps: extracting video metadata, segmenting into initial layers, recursively splitting into deeper layers, and generating thematic collections.}
  \label{fig:pipeline}
\end{figure*}
}

\newcommand{\figureInterface}{
\begin{figure*}[t]
  \includegraphics[width=\textwidth]{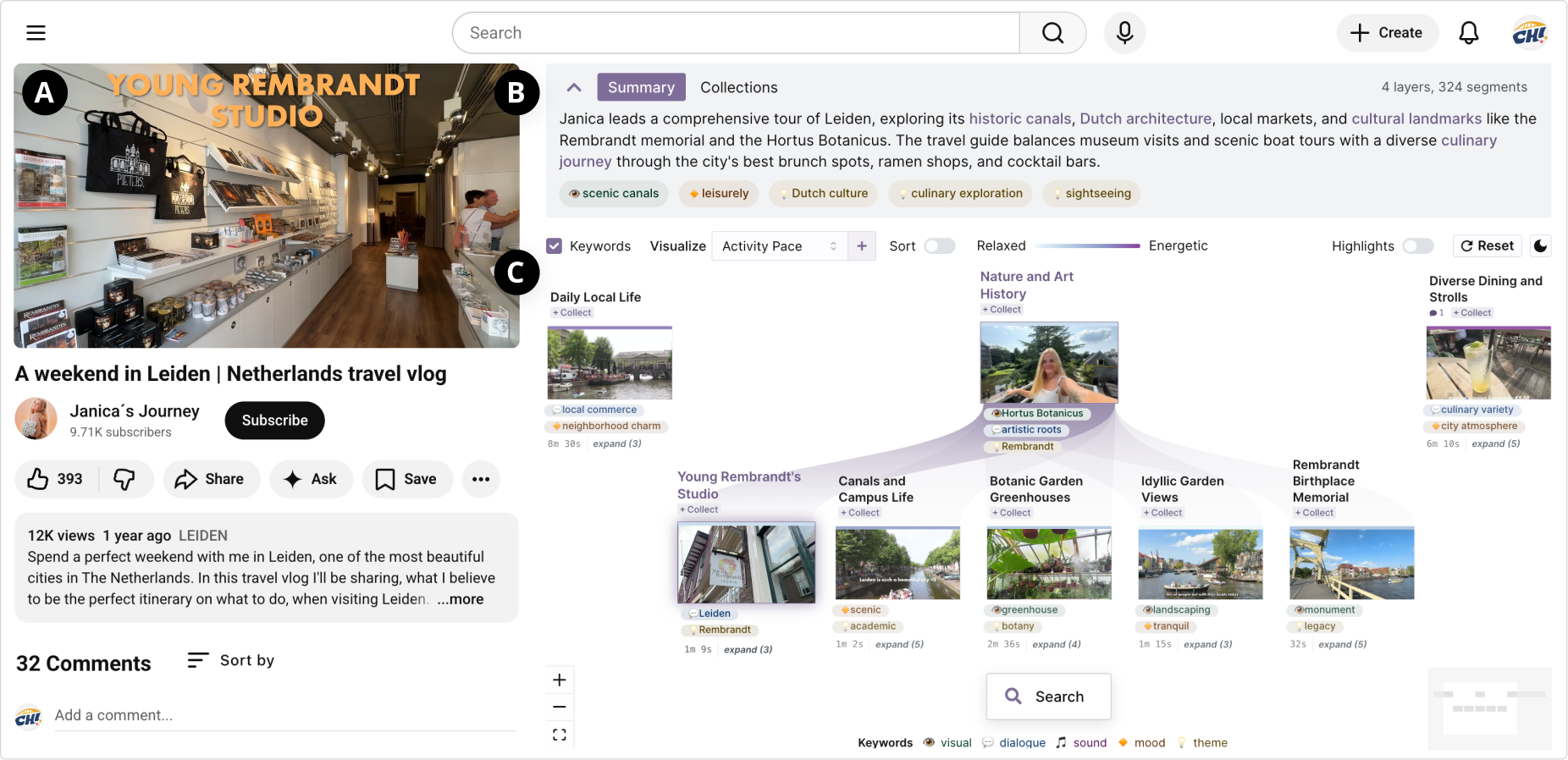}
  \caption{\textbf{Overview of \system's interface. (A)} We keep the standard YouTube video player, description, and comments section on the left, while replacing the right side of the screen with \textbf{(B)} an overall summary and keywords and \textbf{(C)} our hierarchical visualization. The user has expanded a layer 1 node (``Nature and Art History'') to reveal layer 2.
  Video source: \href{https://www.youtube.com/watch?v=57_drtPS9h0}{A weekend in Leiden | Netherlands travel vlog} by \textit{Janica's Journey} (Creative Commons license).}
  \Description{VideoTrail interface integrating video player and hierarchy. The layout places the video player on the left and a summary with keywords and a hierarchical segment map on the right, where a node is expanded to reveal deeper layers.}
  \label{fig:interface}
\end{figure*}
}

\newcommand{\figureAdaptive}{
\begin{figure}[t]
  \includegraphics[width=\linewidth]{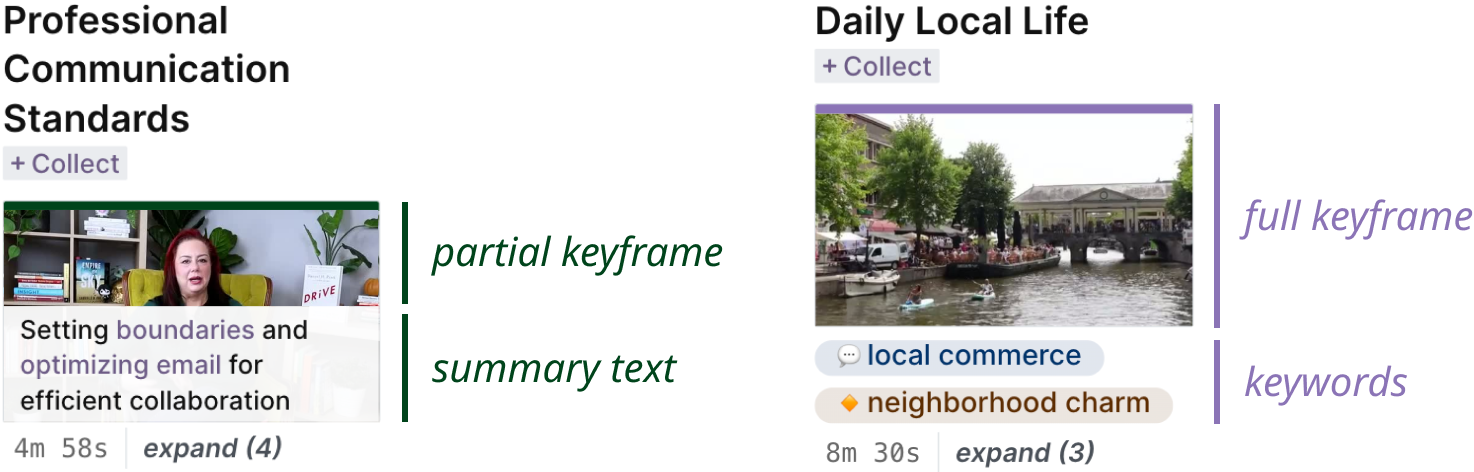}
  \caption{Adaptively weighting textual or visual cues for a \textcolor{DialogueColor}{dialogue-driven} (left) vs. \textcolor{VisualColor}{visual-driven} video (right).}
  \Description{Adaptive display of visual versus textual information cues. For visual-driven videos, keyframes and keywords are emphasized, while for dialogue-driven videos, summaries are prioritized and images are reduced.}
  \label{fig:adaptive}
\end{figure}
}

\newcommand{\figureSearch}{
\begin{figure*}[t]
  \includegraphics[width=\textwidth]{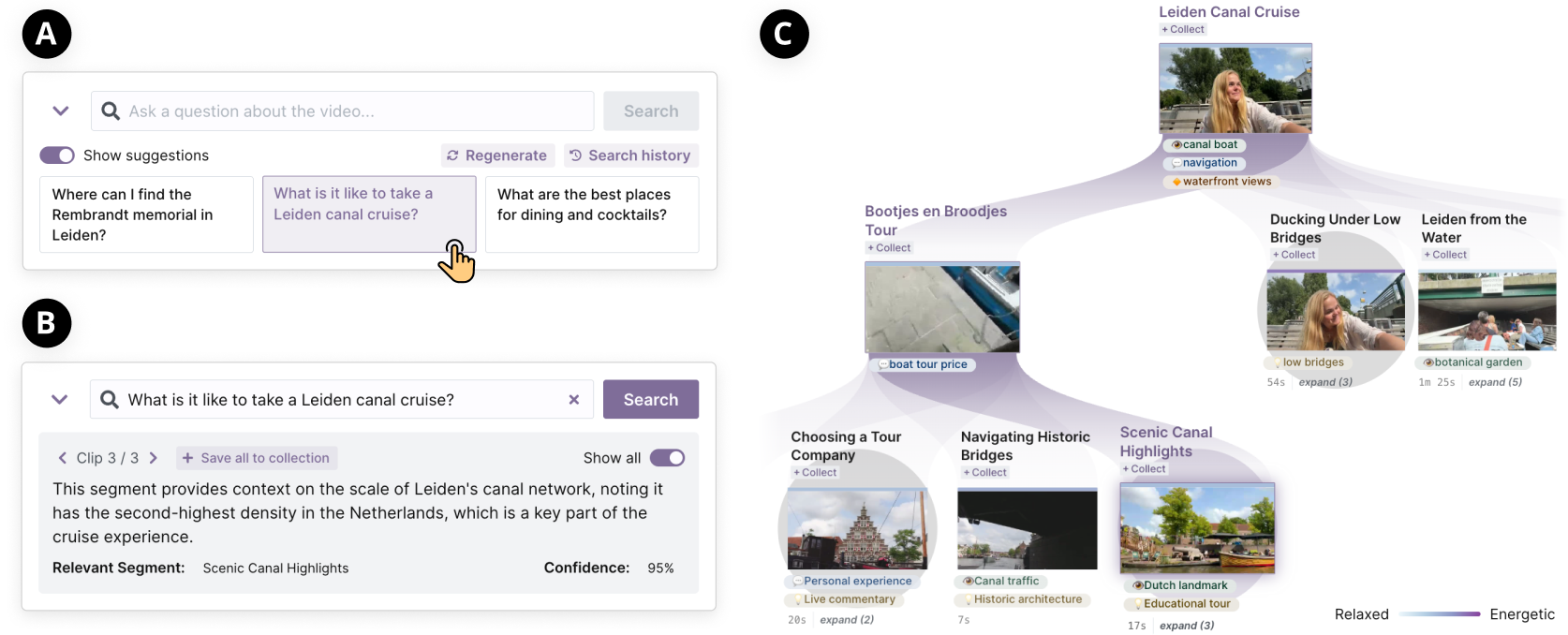}
  \caption{Example of semantic search. (A) Users can expand the search box to ask a question about the video. Question suggestions are provided for inspiration. (B) Search results are displayed in the box when returned and (C) highlighted in the visualization with spotlight circles (\textcolor{PurpleColor}{purple} = current search result, \textcolor{gray}{gray} = other search results). Users can navigate between search results via the search box or by traversing the hierarchy.}
  \Description{Semantic search integrated with hierarchical visualization. A search query returns relevant segments displayed in a list and highlighted within the hierarchy, allowing navigation between results.}
  \label{fig:search}
\end{figure*}
}

\newcommand{\figureLikert}{
\begin{figure*}[t]
  \includegraphics[width=\linewidth]{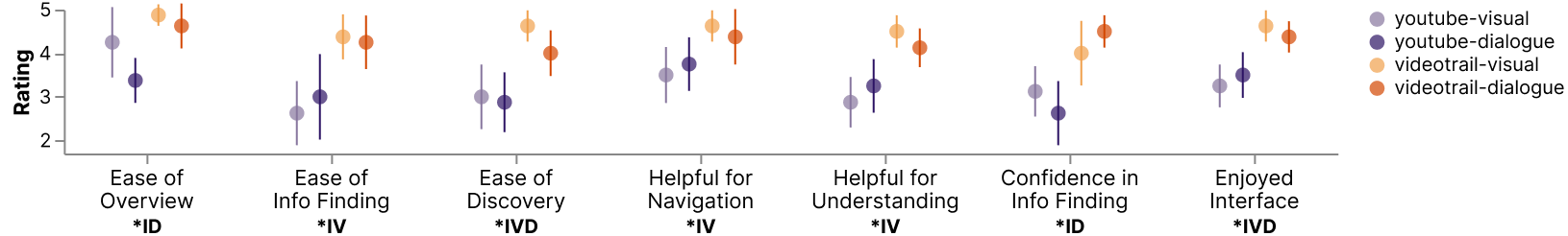}
  \caption{Mean participant ratings (1: strongly disagree - 5: strongly agree) on the utility and ease of use of each interface (\textcolor{PurpleColor}{purple} = YouTube, \textcolor{OrangeColor}{orange} = \system) across video types during the comparison task. Error bars show 95\% CIs. \textbf{*} denotes statistically significant differences ($\mathbf{p < .05}$) between interfaces overall (I) and within a video type (V = visual, D = dialogue).}
  \Description{User ratings comparing VideoTrail and baseline interface. The interval plot shows higher ratings for VideoTrail across all ratings: ease of overview, ease of and confidence in info finding, ease of discovery, helpfulness for navigation, helpfulness for understanding, and interface enjoyment.}
  \label{fig:likert}
\end{figure*}
}

\newcommand{\figureAccTime}{
\begin{figure}[t]
  \includegraphics[width=\linewidth]{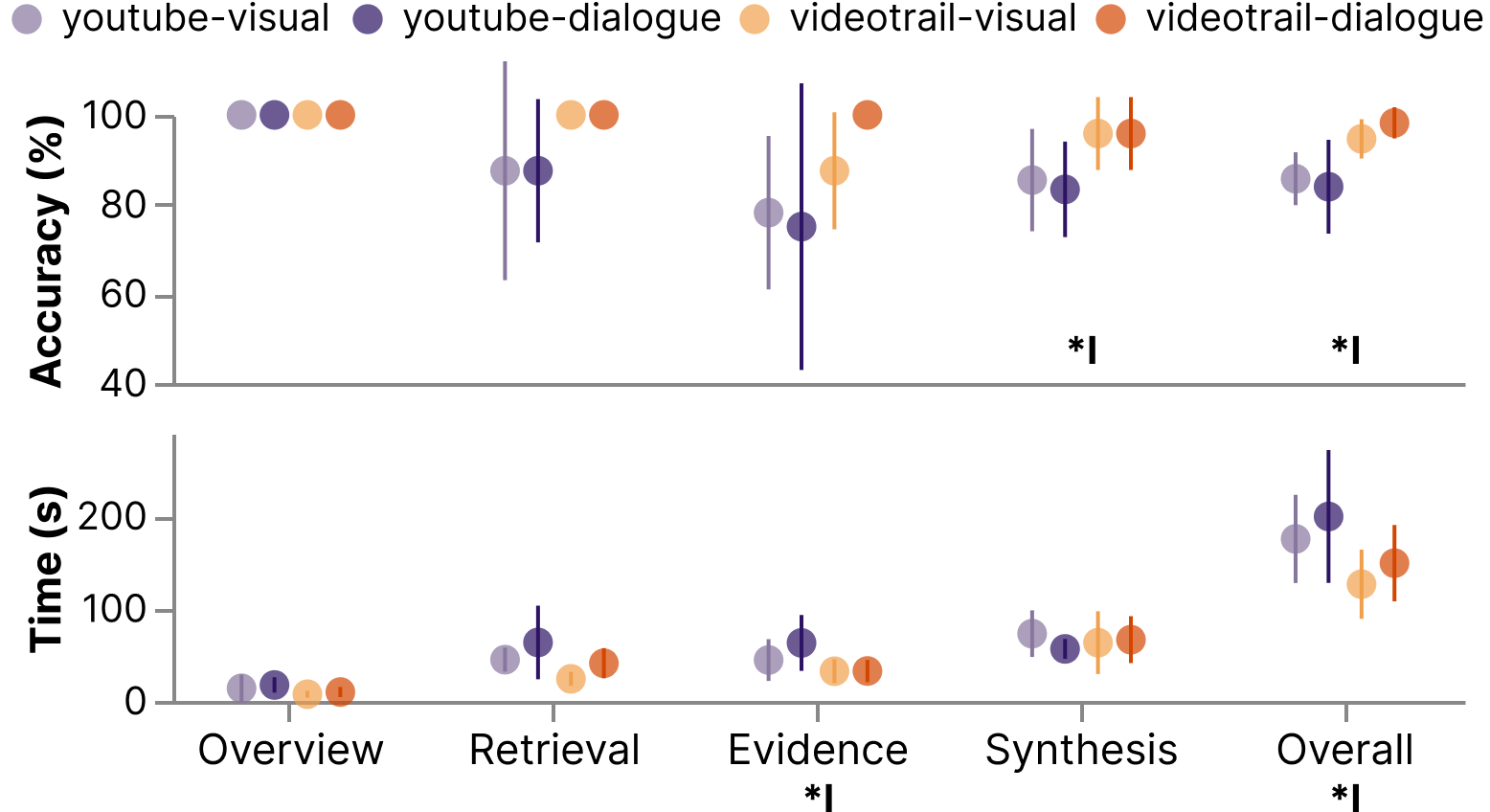}
  \caption{Mean question accuracies (top) and completion times (bottom) with each interface (\textcolor{PurpleColor}{purple} = YouTube, \textcolor{OrangeColor}{orange} = \system) across video types during the comparison task. Error bars show 95\% CIs. \textbf{I*} denotes statistically significant differences ($\mathbf{p < .05}$) between interfaces.}
  \Description{Accuracy and completion time comparison across interfaces. Charts show higher accuracy and generally faster task completion with VideoTrail compared to the baseline interface across task types.}
  \label{fig:acc_times}
\end{figure}
}

\newcommand{\figureExpOld}{
\begin{figure*}[ht]
  \includegraphics[width=\textwidth]{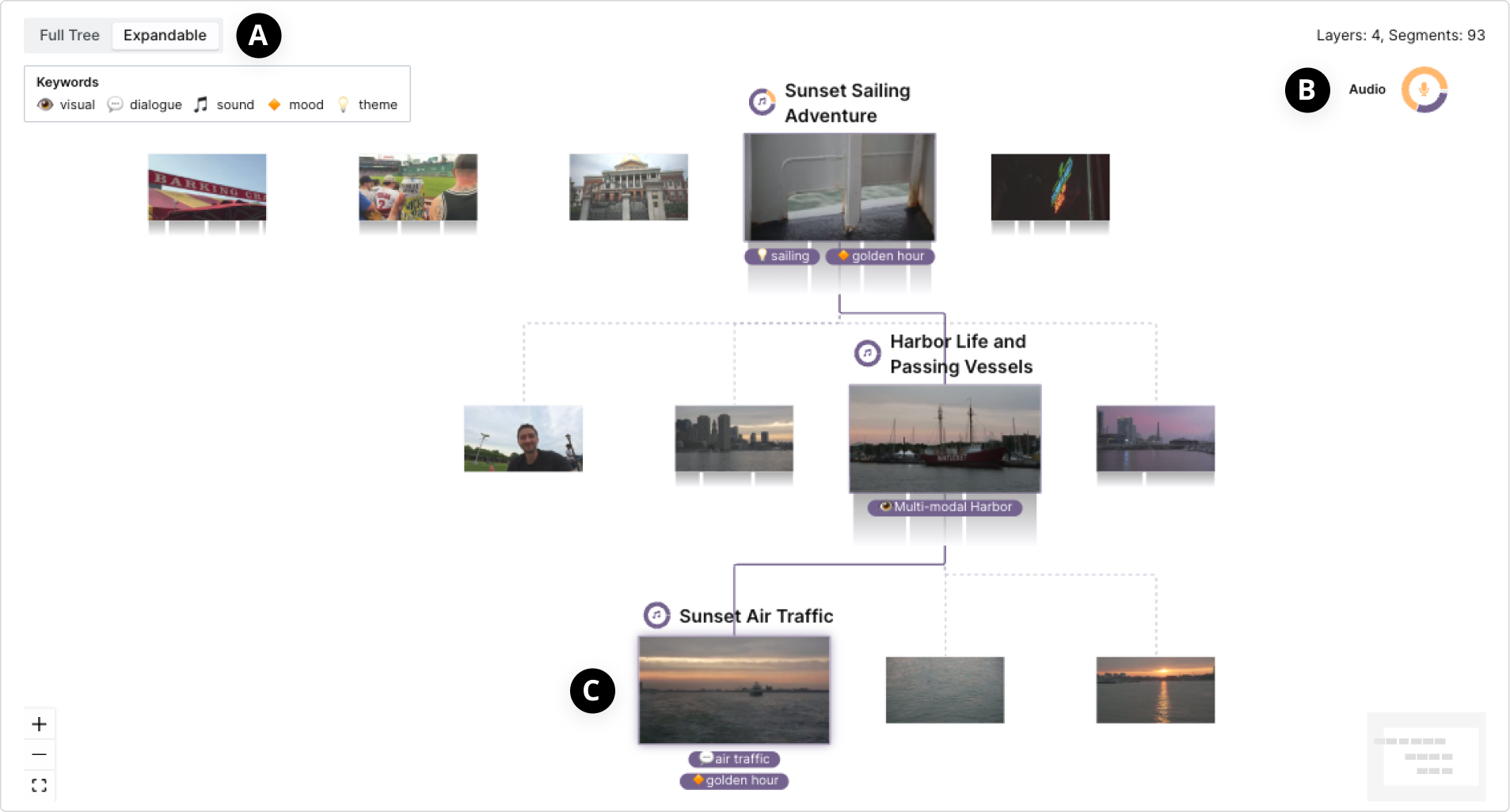}
  \caption{\textbf{Our hierarchical visualization at the time of the formative study,} which included an \textbf{(A)} option to switch between an ``expandable'' and ``full view'' of the video hierarchy. \textbf{(B)} We also visualized the audio distribution of each segment (\eg \textcolor{OrangeColor}{orange} = dialogue, \textcolor{PurpleColor}{purple} = music) and \textbf{(C)} segment styling was different (\eg dashed edges, keyword styling, showing descendants with gray rectangles, scaling when selected).}
  \Description{Early prototype of hierarchical video visualization. The interface shows expandable and full views, audio distribution indicators, and earlier styling of segment nodes and hierarchy connections.}
  \label{fig:exp_old}
\end{figure*}
}

\newcommand{\figureFullOld}{
\begin{figure*}[ht]
  \includegraphics[width=\textwidth]{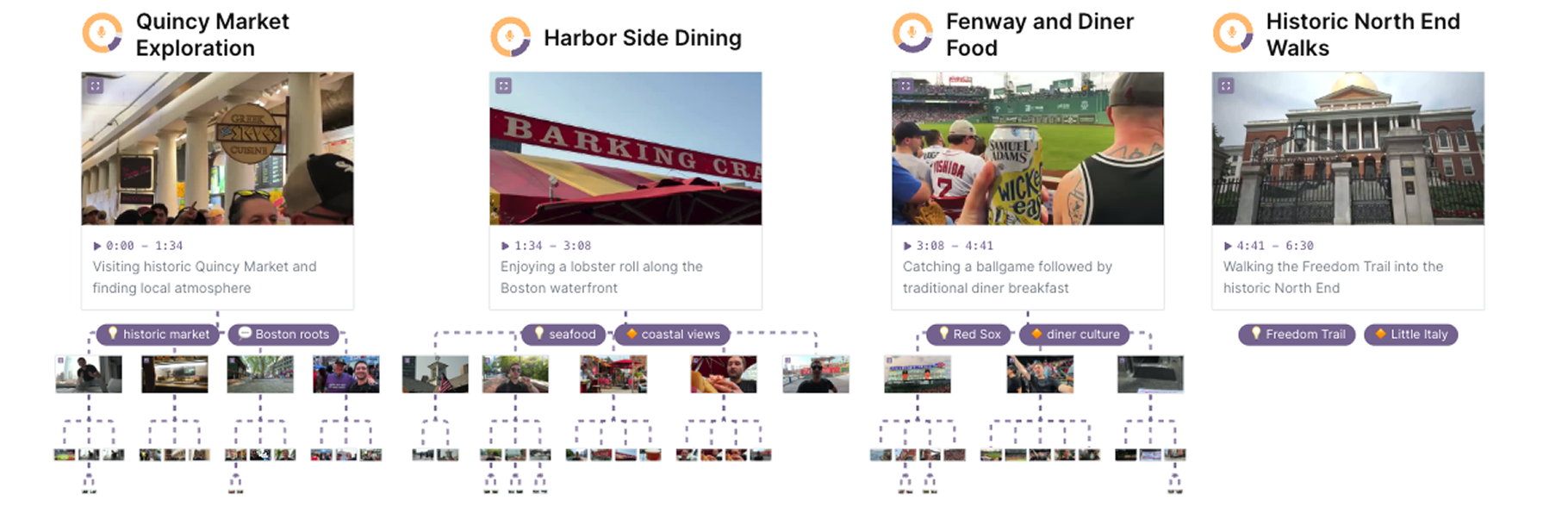}
  \caption{The original ``full view'' of the video hierarchy, which allowed users to view all layers and segments at the same time.}
  \Description{Full hierarchical view of all video segments. The visualization displays all layers and segments simultaneously, showing the complete structure of the video at once.}
  \label{fig:full_old}
\end{figure*}
}

\newcommand{\figureZoom}{
\begin{figure*}[ht]
  \includegraphics[width=0.4\textwidth]{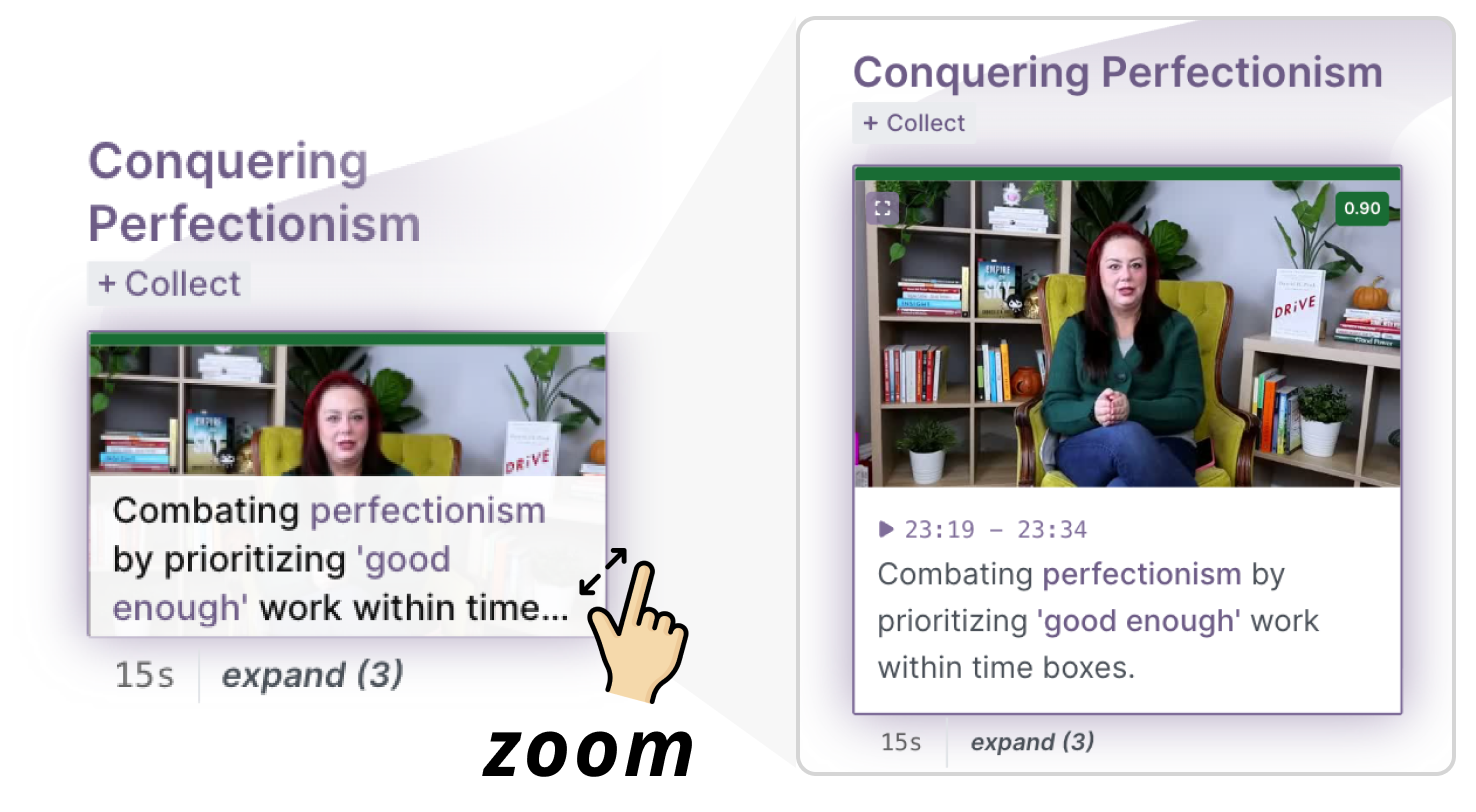}
  \caption{Semantic zoom reveals the full keyframe and summary text for each segment.}
  \Description{Side-by-side comparison of a segment card before and after semantic zoom. The compact Conquering Perfectionism card on the left shows a cropped keyframe and summary, while the enlarged card on the right displays the full keyframe and summary.}
  \label{fig:zoom}
\end{figure*}
}

\newcommand{\figureHighlights}{
\begin{figure*}[ht]
  \includegraphics[width=\textwidth]{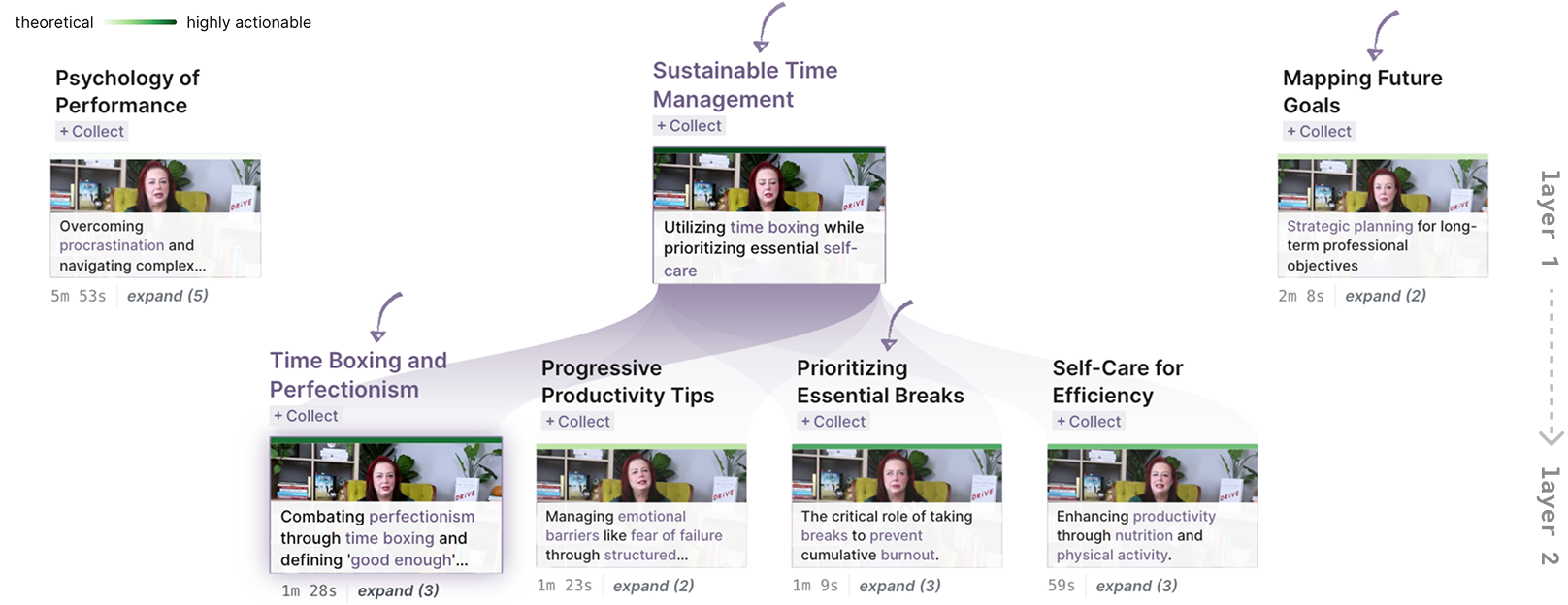}
  \caption{When the ``highlights'' toggle is turned on, purple arrows highlight recommended segments at each layer of the hierarchy. This provides one form of guided exploration of the video.}
  \Description{Highlighted segments for guided exploration. Recommended segments at each layer are marked to guide users through important parts of the video hierarchy.}
  \label{fig:highlights}
\end{figure*}
}

\newcommand{\figureCollections}{
\begin{figure*}[ht]
  \includegraphics[width=\textwidth]{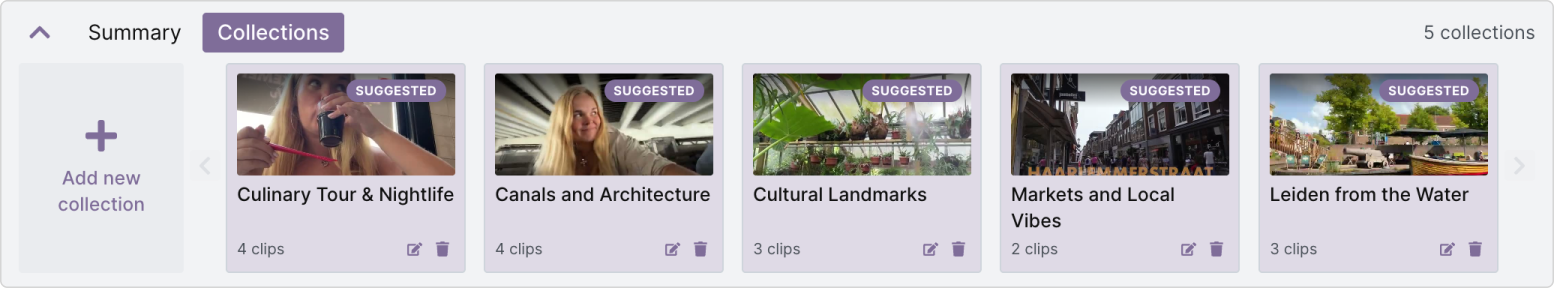}
  \caption{\textbf{\system's collections view}. Each collection groups semantically related segments to highlight themes that may be interesting to viewers, providing another form of guided exploration. Users can also create their own collections.}
  \Description{Collections view grouping related video segments. The interface shows multiple collections of segments organized by theme such as "Culinary Tour & Nightlife" and "Canals and Architecture."}
  \label{fig:collections}
\end{figure*}
}

\newcommand{\figureQualtrics}{
\begin{figure*}[ht]
  \includegraphics[width=\textwidth]{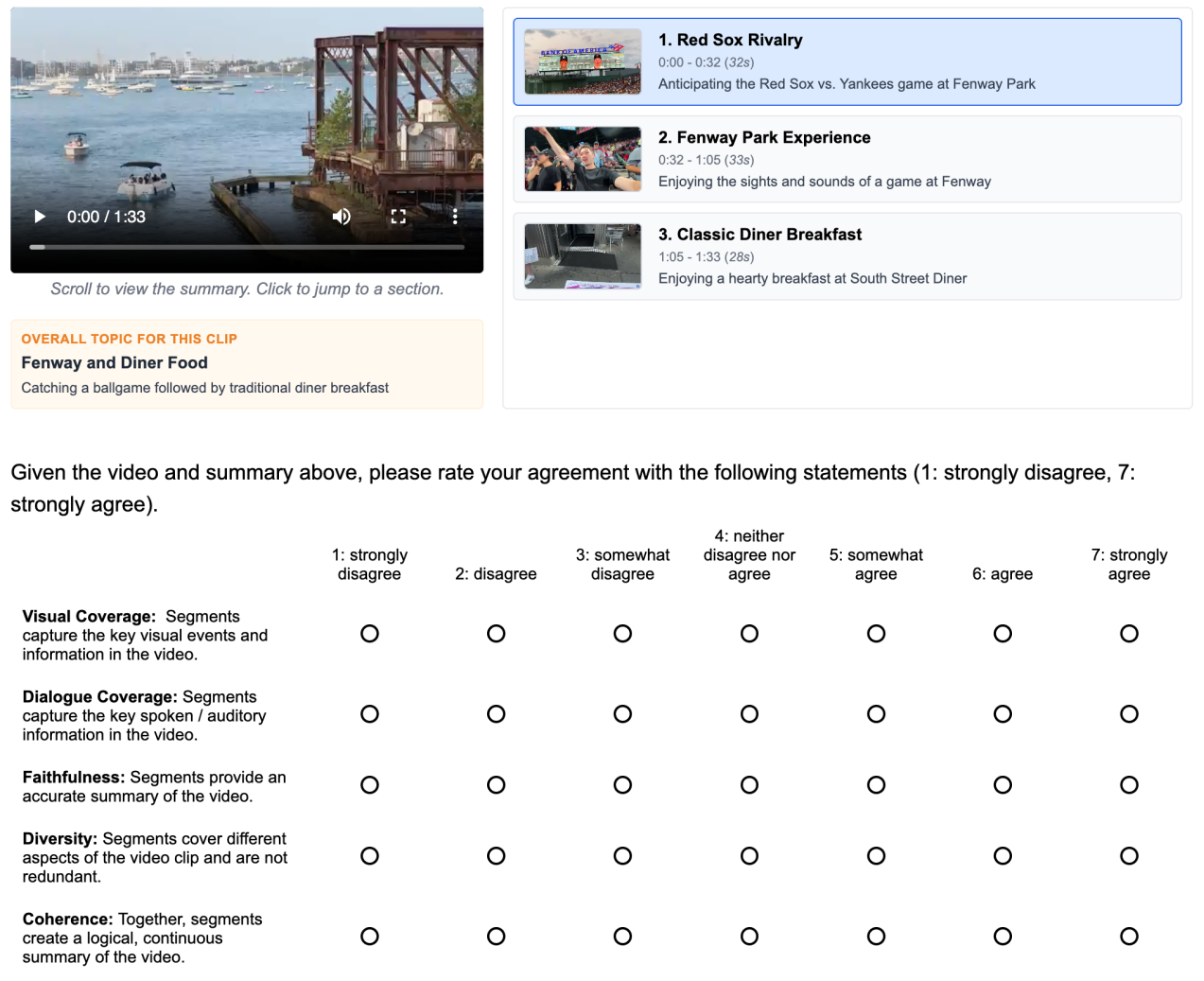}
  \caption{Screenshot of the Qualtrics interface used for evaluating our video segmentations.}
  \Description{Qualtrics rating interface for summary evaluation. A survey screen shows a video clip with one generated summary and Likert-scale questions for judging coverage, faithfulness, boundaries, utility, and overall quality.}
  \label{fig:qualtrics}
\end{figure*}
}

\newcommand{\figureEvalCompare}{
\begin{figure*}[ht]
  \includegraphics[width=\textwidth]{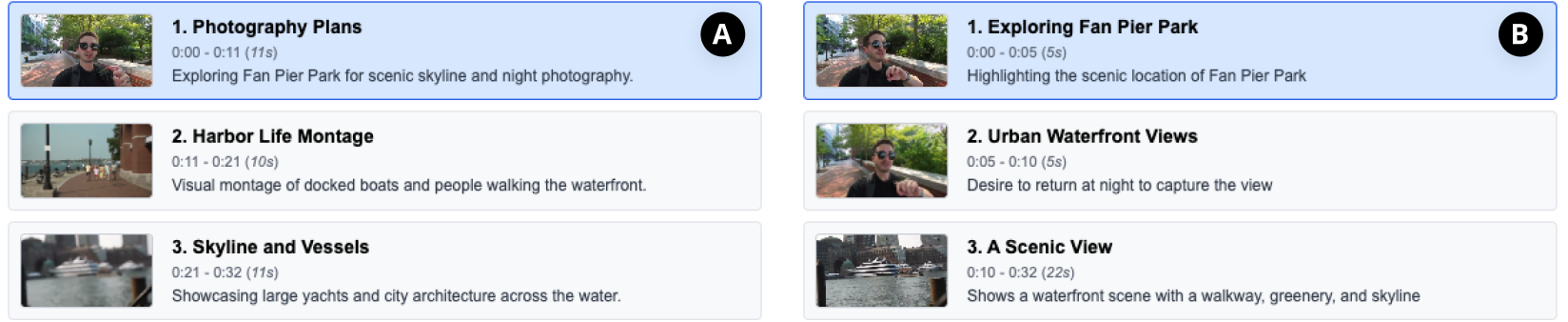}
  \caption{An example summary produced by \textbf{(A)} our pipeline vs. the \textbf{(B)} baseline segmentation model for the same clip.}
  \Description{Side-by-side comparison of two segmentation methods. The figure compares the same video clip summarized by VideoTrail and the baseline, showing differences in segment titles, keyframes, and text descriptions across hierarchical layers.}
  \label{fig:eval_compare}
\end{figure*}
}

\newcommand{\figurePipelineEval}{
\begin{figure*}[ht]
  \includegraphics[width=0.67\textwidth]{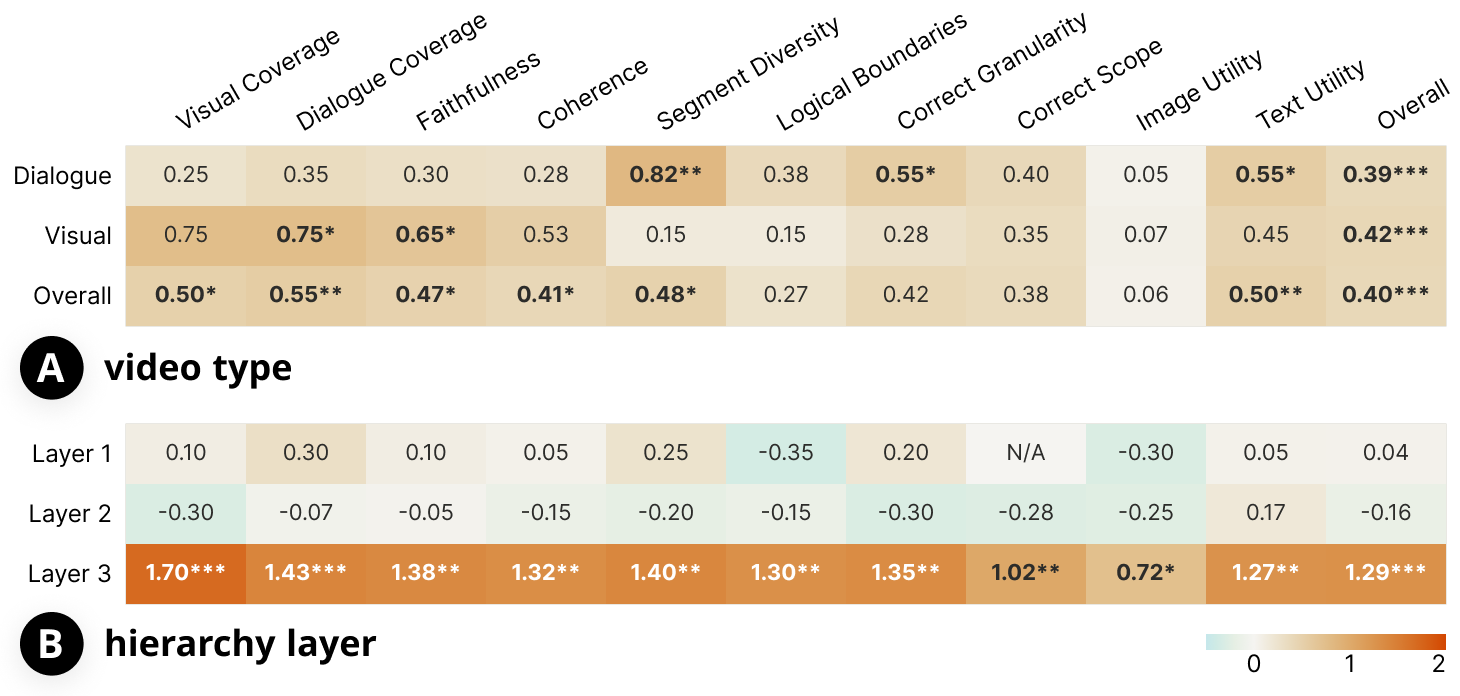}
  \caption{Mean difference in evaluation ratings (our pipeline $-$ baseline)
by \textbf{(A)} video type and \textbf{(B)} hierarchy layer.
Significant results are bolded: $^{*}p < .05$; $^{**}p < .01$; $^{***}p < .001$.}
  \Description{Rating differences between VideoTrail and the baseline. Two plots show mean score differences across evaluation criteria by video type and by hierarchy layer, with larger improvements concentrated at the deepest layer.}
  \label{fig:eval_ratings}
\end{figure*}
}

\newcommand{\tableCompareVids}{
\renewcommand{\arraystretch}{1.2}
\sffamily
\begin{table*}[ht]
{
\begin{tabular}{p{0.2\textwidth} p{0.08\textwidth} p{0.66\textwidth}}
\toprule
\textbf{Video} & \textbf{Category} & \textbf{Questions \& Answers} \\ \midrule

\href{https://www.youtube.com/watch?v=57_drtPS9h0}{A weekend in Leiden | Netherlands travel vlog} by \textit{Janica's Journey} & 
Visual & 
\qa{1. Overview: Provide a 1-2 sentence summary.}{Janica gives a comprehensive tour of Leiden, highlighting its historic canals, cultural landmarks, museums, and local food spots.}
\qa{2. Retrieval: What is the oldest botanical garden in the Netherlands?}{The Hortus Botanicus (15:51).}
\qa{3. Evidence: What makes Paco Ciao, the first brunch spot the narrator visits, unique?}{Its entrance is hidden through a closet (1:41).

\textit{Alternate answers:} It has unique menu items like pink pancakes with fairy floss (3:05), and unique decor like a disco bathroom (3:21). 
}
\qa{4. Synthesis: How do the two Rembrandt sites in Leiden differ?}{Rembrandt's studio is a free site for his early life history (13:49), while his childhood home is now replaced by a memorial because the original site was demolished (13:55–14:11; 19:29–19:43).}
  \\ \midrule

\href{https://www.youtube.com/watch?v=w7eK-5SCzmo}{How to make the best Homemade Vanilla cake} by \textit{MR. NYSauce} & 
Visual & 
\qa{1. Overview: Provide a 1-2 sentence summary.}{This tutorial demonstrates the step-by-step process of baking a moist vanilla cake from scratch and preparing buttercream frosting for final assembly.}
\qa{2. Retrieval: How much powdered sugar and heavy whipping cream are needed for the frosting?}{4.5 cups of powdered sugar and 4 tablespoons of whipping cream (14:30).} 
\qa{3. Evidence: What is the desired consistency of the butter for the frosting, and what happens if it's too soft?}{It should be soft to the touch but not too soft (14:40-14:46). If it's too soft, the mix will be very loose  (14:46-15:00).} 
\qa{4. Synthesis: What is a difference in how the cake batter and frosting are mixed, for example, in terms of technique or desired consistency?}{The cake should be a nice and thick batter (11:29). 
For the frosting, a thicker, stiff consistency is desired so it does not fall off the spoon and holds its shape when twisted (24:09).

\textit{Alternate answer:} For the cake, the flour mixture and buttermilk are alternated in the mixer to create the final batter (7:30).
For the frosting, the heavy whipping cream is added as the last step to achieve the desired consistency (22:23).
} 
\\ \midrule

\href{https://www.youtube.com/watch?v=W3RLUa4nJas}{How to Finally Stop Struggling with Productivity} by \textit{She Makes It Work} & 
Dialogue & 
\qa{1. Overview: Provide a 1-2 sentence summary.}{This guide explores practical strategies for boosting productivity by managing information overload, setting clear boundaries, and overcoming emotional barriers like perfectionism.}
\qa{2. Retrieval: What is the one-minute email rule?}{The rule suggests that if an email is going to take less than a minute to deal with, you should deal with it right away (7:02-7:04).} 
\qa{3. Evidence: How does the speaker argue that taking breaks ultimately helps you ``get time back''?}{Taking breaks helps reset your brain, allowing you to work more efficiently. Putting off breaks decreases your work quality, requiring more time to make up for it (27:28-27:36).} 
\qa{4. Synthesis: How do the ideas "done is better than perfect" and ``timeboxing'' relate?}{``Done is better than perfect'' encourages people to prioritize completion over perfectionism (21:18-21:26). ``Timeboxing'' then provides a practical framework for this by allocating a fixed amount of time to a task (23:25-23:30).} 
\end{tabular}
}
\end{table*}

\begin{table*}[ht]
{
\begin{tabular}{p{0.2\textwidth} p{0.08\textwidth} p{0.66\textwidth}}
\toprule
\textbf{Video} & \textbf{Category} & \textbf{Questions \& Answers}
\\ \midrule

\href{https://www.youtube.com/watch?v=GVHyV9zhQP4}{Planner Haul Review ft. Happiness Planner, Erin Condren, Get To Work, Day Designer and More!} by \textit{christopher allen} & 
Dialogue & 
\qa{1. Overview: Provide a 1-2 sentence summary.}{This review video evaluates eight distinct planning systems, comparing features, aesthetics, and functionality to help viewers find an ideal organizational tool for their needs.}
\qa{2. Retrieval: What is the price of the TF Publishing Palm Planner, and who is it recommended for?}{The planner is under \$10 and recommended for students or anyone new to planning (2:38).} 
\qa{3. Evidence: What is the narrator's main critique of the Day Designer planner?}{The flimsy paper cover on the back makes it prone to bending and creasing (14:55).

\textit{Alternate answer:} He also dislikes the consistent lavender color scheme throughout the planner (16:13-16:22).} 
\qa{4. Synthesis: How does the Get To Work planner differ from the Happiness planner?}{The Get To Work planner is designed for daily task management, with a minimalist layout and high durability (19:46).
The Happiness Planner is a specialized journal for tracking happiness and personal growth, and comes with a companion app (25:55-26:16).
} 
\\ 
\end{tabular}
}
\caption{Overview of the four videos, questions, and reference answers we used for the comparison task. All selected videos have a Creative Commons license.}
\Description{Questions and answers used in the comparison task. The table lists four videos with overview, retrieval, evidence, and synthesis questions, along with reference answers and timestamps for scoring.}
\label{tab:comp_vids}
\end{table*}
}

\newcommand{\tableExploreVids}{
\renewcommand{\arraystretch}{1.2}
\sffamily
\begin{table*}[ht]
{
\begin{tabular}{p{0.04\textwidth} p{0.45\textwidth} p{0.45\textwidth}}
\toprule
\textbf{ID} & \textbf{Familiar Video \& Goal} & \textbf{Unfamiliar Video \& Goal} \\ \midrule

P1 & 
\href{https://www.youtube.com/watch?v=hNCnZ0H34NQ}{Toronto Travel Guide - Complete City Tour - The Best Attractions, Transport, Sports \& More (Canada)} by \textit{Ultimate Bucket List} 

\textbf{Goal:} Plan summer travel to attractions around the city, in particular the downtown area.
& 
\href{https://www.youtube.com/watch?v=XstMdaIqjVA}{The Entire History of the Middle East} by \textit{This Is History} 

\textbf{Goal:} Learn more about the history of the Middle East, especially the Persian empire.
\\ \midrule

P2 & 
\href{https://www.youtube.com/watch?v=QP8I9YIjGc8}{Why are we even doing this..} by \textit{Hasan Abi} 

\textbf{Goal:} Remind myself of the divergent interests between the US and Israel.
& 
\href{https://www.youtube.com/watch?v=YFjfBk8HI5o}{OpenClaw: The Viral AI Agent that Broke the Internet - Peter Steinberger | Lex Fridman Podcast \#491} by \textit{Lex Friedman} 

\textbf{Goal:} Learn how Peter does the ``agentic engineering'' to build all these tools.
\\ \midrule

P3 & 
\href{https://www.youtube.com/watch?v=7bjYAnSLBZo}{
I Tried The Best Convenience Stores in America} by \textit{Safiya Nygaard} 

\textbf{Goal:} Familiarize myself with the best convenience stores for a road trip.
& 
\href{https://www.youtube.com/watch?v=ekWc5TlNjK4}{First Class on the World's Newest Airline (Global Airlines)} by \textit{Kara and Nate} 

\textbf{Goal:} Learn about this new airline and the amenities that they provide.
\\ \midrule

P4 & 
\href{https://www.youtube.com/watch?v=x32Zq-XvID4}{The Obviously True Theorem No One Can Prove} by \textit{Veritasium
 and fern} 

\textbf{Goal:} Relearn about Goldbach's Conjecture.
& 
\href{https://www.youtube.com/watch?v=YQGai2PVHBs}{How 20 Cheeses Are Made Around The World | Regional Eats | Insider Food Marathon} by \textit{Insider Food} 

\textbf{Goal:} Learn more about gastronomy studies and different cheeses.
\end{tabular}
}
\end{table*}

\begin{table*}
{
\begin{tabular}{p{0.04\textwidth} p{0.45\textwidth} p{0.45\textwidth}}
\toprule
\textbf{ID} & \textbf{Familiar Video \& Goal} & \textbf{Unfamiliar Video \& Goal}
\\ \midrule

P5 & 
\href{https://www.youtube.com/watch?v=gh7tff8roKw}{Understanding and Steering Generative AI Systems} by \textit{Simons Institute for the Theory of Computing} 

\textbf{Goal:} Refresh on ideas from this talk, like TextSpan.
& 
\href{https://www.youtube.com/watch?v=-Mvyh66WI8U}{Automating the US Welfare State: Historical Experiments in Technological Federalism} by \textit{Simons Institute for the Theory of Computing} 

\textbf{Goal:} Get key insights from this talk that I haven't gotten around to watching yet.
\\ \midrule

P6 & 
\href{https://www.youtube.com/watch?v=kNNFcY6XFLg}{
My First DNF at the Mesa Marathon | Full Breakdown of What Happened} by \textit{Nico Felich} 

\textbf{Goal:} Revisit the reflections about running in this video.
& 
\href{https://youtu.be/XSzpQDQ7r4U?si=z42ViwLXSn4AVEb2}{
Extended interview: Jack Schlossberg} by \textit{CBS Sunday Morning} 

\textbf{Goal:} Learn more about Jack Schlossberg's campaign as I've been following him for a while.
\\ \midrule

P7 & 
\href{https://www.youtube.com/watch?v=Pp2wbyLoEtM}{Fear of Cold} by \textit{Jacob Geller} 

\textbf{Goal:} Revisit this to analyze how he blended different types of media to tell a very interesting story.
& 
\href{https://www.youtube.com/watch?v=x-XdOaZPhBw}{
Why This Island Could Trigger World War 3} by \textit{Johnny Harris} 

\textbf{Goal:} Learn more about the state of the world and Taiwan becoming a global semiconductor hub.
\\ \midrule

P8 & 
\href{https://www.youtube.com/watch?v=zTXfu1GUPM0}{The Protestant Reformation - Church History Simplified} by \textit{Redeemed Zoomer} 

\textbf{Goal:} Relearn facts about the Protestant Reformation.
& 
\href{https://www.youtube.com/watch?v=BX7AhPb2aj8}{
51 Tips I Wish I Knew Before Visiting Tokyo, Japan} by \textit{Camden David} 

\textbf{Goal:} Plan future travel to Japan.
\\ \midrule

P9 & 
\href{https://www.youtube.com/watch?v=LLoO24HLr1o}{
I Took the Longest Bus Ride in America} by \textit{Ryan Trahan} 

\textbf{Goal:} Revisit this video to learn what traveling across America on a bus looks like.
& 
\href{https://www.youtube.com/watch?v=5I1ZmW_iPAw}{
\$50. 3 Meals. 1 Day. BOSTON | Feasting on \$50} by \textit{Martinez \& Mu} 

\textbf{Goal:} Learn about different food places in Boston to try.
\\ \midrule

P10 & 
\href{https://www.youtube.com/watch?v=i3aK-EK5V2k}{
Why the Music in Cats (2019) is Worse than you Thought} by \textit{Sideways} 

\textbf{Goal:} Explore why playing the old music in the soundtrack detracts from the soul of the new one.
& 
\href{https://www.youtube.com/watch?v=-vNp8tXQdow}{
Why ICE And CBP Are Unreformable, With Spencer Ackerman | The Intercept Briefing} by \textit{The Intercept} 

\textbf{Goal:} Learn more about our current wars ``overseas'' and how it ties into what ICE and CBP are doing.
\\ \midrule

P11 & 
\href{https://www.youtube.com/watch?v=P90-6W67564}{
Abaqus Heat Transfer Analysis 1 | Steady State Conduction through a Square Plate} by \textit{A-Star Education} 

\textbf{Goal:} Revisit to ensure I'm using the software correctly to build a model. 
& 
\href{https://www.youtube.com/watch?v=MNbkNmNV9J4}{
Getting Started with Rhino Part 2 - Lines, Faces, and 3D Shapes!} by \textit{The Rhino Essentials} 

\textbf{Goal:} Learn this new software for my future possible drawing.
\\ \midrule

P12 & 
\href{https://www.youtube.com/watch?v=d8PndpFPL8g}{
Populism, Media Revolutions, and Our Terrible Moment} by \textit{vlogbrothers} 

\textbf{Goal:} Understand how these ideas can be applied to the modern AI era.
& 
\href{https://www.youtube.com/watch?v=VU7lh-BUoEQ}{
The Man Who Solved Addiction | Carl Jung} by \textit{Aperture} 

\textbf{Goal:} Learn more about the psychology of addiction.
\\ \midrule

P13 & 
\href{https://www.youtube.com/watch?v=zeJD6dqJ5lo}{But what is the Central Limit Theorem?} by \textit{3Blue1Brown} 

\textbf{Goal:} Review my basic understanding of stats concepts to apply to work in my current classes.
& 
\href{https://www.youtube.com/watch?v=U6JYek1HBJ0}{
making clay trinkets using air dry clay using air dry clay / no bake} by \textit{gazoo studio} 

\textbf{Goal:} Learn how to make little clay keychains and trinkets as gifts for my friends and family.
\\ \midrule

P14 & 
\href{https://www.youtube.com/watch?v=BN-OW2oNKRg}{
The Future of Robotics Hype, Breakthroughs, and Challenges with Professor Russ Tedrake} by \textit{MIT CSAIL Alliances} 

\textbf{Goal:} Understand the main rationale of the professor being interviewed on the future of robotics.
& 
\href{https://www.youtube.com/watch?v=5Nkad_6aigM}{
The Impossible Landing} by \textit{neo} 

\textbf{Goal:} Learn more about the ``impossible landing'' and historical aviation events.
\\ \midrule

P15 & 
\href{https://www.youtube.com/watch?v=-JRSF-zDgvk}{
Make Beautiful Sourdough With Claire Saffitz | Try This at Home | NYT Cooking} by \textit{NYT Cooking} 

\textbf{Goal:} Review steps on making sourdough bread.
& 
\href{https://www.youtube.com/watch?v=AaVJX_GevpI}{
5 EASY Watercolour Flowers - Perfect For Beginners! (it's easier than you think)} by \textit{Emma Jane Lefebvre (EmJ Watercolour Studio)} 

\textbf{Goal:} Learn how to paint watercolor paintings because I just bought some supplies.
\\ \midrule

P16 & 
\href{https://www.youtube.com/watch?v=wO61D9x6lNY}{Outside In} by \textit{ssgelm} 

\textbf{Goal:} Revisit now that I know a lot more algebraic topology and view it from a more rigorous mathematical perspective. 
& 
\href{https://www.youtube.com/watch?v=p_di4Zn4wz4}{Differential equations, a tourist's guide | DE1} by \textit{3Blue1Brown} 

\textbf{Goal:} Learn about differential equations, as they're ubiquitous and foundational. 
\end{tabular}
}
\caption{Overview of the 32 videos (16 familiar and 16 unfamiliar) videos, and their corresponding information goals, selected by participants for the exploration task.}
\Description{Participant-selected videos and exploration goals. The table summarizes sixteen participants’ chosen familiar and unfamiliar videos with corresponding information-seeking goals across diverse topics.}
\label{tab:exp_vids}
\end{table*}
}

\newcommand{\tableCompareResults}{
\begin{table*}[ht]
\caption{%
  Linear mixed model (LMM) results for the interface effect per question, with exploratory
  simple-effects Welch $t$-tests within each video type (YT\,=\,YouTube, VT\,=\,\system).
  LMM: \texttt{rating \textasciitilde{} interface + video\_type} with random intercepts per
  participant, fitted via maximum likelihood.
  All $p$-values are BH-corrected across questions within each video type.
  Means are on a 1--5 Likert scale.
  $\mathbf{^{*}p < .05}$; $\mathbf{^{**}p < .01}$; $\mathbf{^{***}p < .001}$.
  Statistically significant results are bolded.
}
\Description{Participant ratings across interfaces and video types. The table reports mean ratings and statistical test results for each usability measure, comparing VideoTrail and baseline across visual and dialogue-based videos.}
\label{tab:likert_llm}
\sffamily
\small
\renewcommand{\arraystretch}{1.2}
\begin{tabular}{p{0.22\textwidth} p{0.2\textwidth} p{0.25\textwidth} p{0.25\textwidth}}
\toprule
\textbf{Question}
  & \textbf{Overall (LMM)}
  & \textbf{Visual Videos}
  & \textbf{Dialogue Videos} \\
\midrule
 
Ease of Overview
  & $\beta = 0.94$, $\mathbf{p < .001^{***}}$
  & YT: $M = 4.25$, VT: $M = 4.88$;\newline $t(8.3) = -1.45$, $p = .183$
  & YT: $M = 3.38$, VT: $M = 4.63$;\newline $t(14.0) = -3.36$, $\mathbf{p = .016^{*}}$ \\ \midrule
 
Ease of Info Finding
  & $\beta = 1.50$, $\mathbf{p < .001^{***}}$
  & YT: $M = 2.63$, VT: $M = 4.38$;\newline $t(12.5) = -3.82$, $\mathbf{p = .005^{**}}$
  & YT: $M = 3.00$, VT: $M = 4.25$;\newline $t(11.8) = -2.12$, $p = .066$ \\ \midrule
 
Ease of Discovery
  & $\beta = 1.38$, $\mathbf{p < .001^{***}}$
  & YT: $M = 3.00$, VT: $M = 4.63$;\newline $t(10.1) = -3.87$, $\mathbf{p = .005^{**}}$
  & YT: $M = 2.88$, VT: $M = 4.00$;\newline $t(13.1) = -2.55$, $\mathbf{p = .042^{*}}$ \\ \midrule
 
Helpful for Navigation
  & $\beta = 0.88$, $\mathbf{p < .001^{***}}$
  & YT: $M = 3.50$, VT: $M = 4.63$;\newline $t(11.0) = -3.00$, $\mathbf{p = .017^{*}}$
  & YT: $M = 3.75$, VT: $M = 4.38$;\newline $t(14.0) = -1.39$, $p = .187$ \\ \midrule
 
Helpful for Understanding
  & $\beta = 1.25$, $\mathbf{p < .001^{***}}$
  & YT: $M = 2.88$, VT: $M = 4.50$;\newline $t(11.9) = -4.64$, $\mathbf{p = .002^{**}}$
  & YT: $M = 3.25$, VT: $M = 4.13$;\newline $t(12.7) = -2.26$, $p = .059$ \\ \midrule
 
Confidence in Info Finding
  & $\beta = 1.38$, $\mathbf{p < .001^{***}}$
  & YT: $M = 3.13$, VT: $M = 4.00$;\newline $t(13.2) = -1.83$, $p = .106$
  & YT: $M = 2.63$, VT: $M = 4.50$;\newline $t(10.3) = -4.47$, $\mathbf{p = .008^{**}}$ \\ \midrule
 
Enjoyed Interface
  & $\beta = 1.13$, $\mathbf{p < .001^{***}}$
  & YT: $M = 3.25$, VT: $M = 4.63$;\newline $t(12.8) = -4.44$, $\mathbf{p = .002^{**}}$
  & YT: $M = 3.50$, VT: $M = 4.38$;\newline $t(12.4) = -2.70$, $\mathbf{p = .042^{*}}$ \\

\bottomrule
 
\end{tabular}
\end{table*}
}

\newcommand{\tableScoreResults}{
\begin{table*}[ht]
\caption{%
  LMM results for the interface effect on accuracy (\textbf{a}) and completion time (\textbf{b})
  per question (YT\,=\,YouTube, VT\,=\,\system).
  Model: \texttt{value \textasciitilde{} interface + video\_type} with random intercepts per
  participant, fitted via maximum likelihood; $p$-values BH-corrected across questions.
  Overview is excluded from accuracy as all scores were 100\%.
  Negative $\beta$ in (b) indicates faster completion with \system.
  $\mathbf{^{*}p < .05}$; $\mathbf{^{**}p < .01}$; $\mathbf{^{***}p < .001}$.
  Statistically significant results are bolded.
}
\Description{Accuracy results and completion times across interfaces and tasks. The table shows statistical comparisons between VideoTrail and baseline across overview, retrieval, evidence, and synthesis tasks.}
\sffamily
\small
\renewcommand{\arraystretch}{1.2}
\begin{minipage}[t]{0.48\textwidth}
\centering
\textbf{(a) Accuracy}\\[4pt]
\begin{tabular}{p{0.25\textwidth} p{0.38\textwidth}}
\toprule
\textbf{Question} & \textbf{LMM ($\beta_{\text{interface}}$, $p$)} \\
\midrule
Overview  & ---\\
Retrieval  & $\beta = 12.50$, $p = .074$\\
Evidence   & $\beta = 17.19$, $p = .074$ \\
Synthesis  & $\beta = 11.46$, $\mathbf{p = .010^{*}}$ \\
Overall    & $\beta = 11.61$, $\mathbf{p < .001^{***}}$ \\

\bottomrule
\end{tabular}
\end{minipage}
\hfill
\begin{minipage}[t]{0.48\textwidth}
\centering
\textbf{(b) Completion Time}\\[4pt]
\begin{tabular}{p{0.25\textwidth} p{0.38\textwidth}}
\toprule
\textbf{Question} & \textbf{LMM ($\beta_{\text{interface}}$, $p$)} \\
\midrule
Overview   & $\beta = -7.00$\,s, $p = .108$ \\
Retrieval & $\beta = -21.63$\,s, $p = .056$\\
Evidence  & $\beta = -21.25$\,s, $\mathbf{p = .045^{*}}$ \\
Synthesis & $\beta = 0.13$\,s, $p = .991$ \\
Overall   & $\beta = -49.75$\,s, $\mathbf{p = .035^{*}}$ \\

\bottomrule
\end{tabular}
\end{minipage}
\label{tab:scores}
\end{table*}
}

\newcommand{\tableVideoType}{%
\begin{table*}[ht]
\caption{%
  Wilcoxon signed-rank tests comparing \system (VT) vs.\ baseline (B) ratings
  (1--7 scale), stratified by video type and question.
  Each cell reports the paired means and test result:
  B\,=\,baseline mean, VT\,=\,\system mean, $\Delta$\,=\,VT\,$-$\,B,
  $W$\,=\,Wilcoxon statistic, $p$\,=\,two-tailed $p$-value.
  The \emph{Overall} column pools all questions within each row;
  the \emph{Overall} row pools all video types within each column.
  Significant results ($p < .05$) are \textbf{bolded}.
  $^{*}p < .05$;\;$^{**}p < .01$;\;$^{***}p < .001$.
}
\Description{Comparison of VideoTrail and baseline ratings by video type. The table reports paired mean ratings and Wilcoxon signed-rank test results for each evaluation criterion, separated by dialogue-driven and visual-driven videos.}
\label{tab:vt_vs_baseline_videotype}
\sffamily
\renewcommand{\arraystretch}{1.3}
\setlength{\tabcolsep}{4pt}
\scriptsize
\resizebox{\textwidth}{!}{%
\begin{tabular}{p{0.05\textwidth}
                p{0.07\textwidth} p{0.07\textwidth} p{0.07\textwidth}
                p{0.07\textwidth} p{0.07\textwidth} p{0.07\textwidth}
                p{0.07\textwidth} p{0.07\textwidth} p{0.07\textwidth}
                p{0.07\textwidth} p{0.07\textwidth}}
\toprule
\thead{Video \\ Type} &
  \thead{Visual \\ Coverage} &
  \thead{Dialogue \\ Coverage} &
  \thead{Faithfulness} &
  \thead{Coherence} &
  \thead{Segment \\ Diversity} &
  \thead{Logical \\ Boundaries} &
  \thead{Correct \\ Granularity} &
  \thead{Correct \\ Scope} &
  \thead{Image \\ Utility} &
  \thead{Text \\ Utility} &
  \thead{Overall} \\
\midrule
 
\textbf{Dialogue}
  & \ns{\cell{5.07}{5.32}{0.25}{103.0}{$p\!=\!.285$}}
  & \ns{\cell{5.22}{5.57}{0.35}{69.5}{$p\!=\!.107$}}
  & \ns{\cell{5.38}{5.68}{0.30}{58.0}{$p\!=\!.133$}}
  & \ns{\cell{5.37}{5.65}{0.28}{59.0}{$p\!=\!.144$}}
  & \sig{\cell{4.87}{5.68}{0.82}{38.5}{$\mathbf{p\!=\!.007^{**}}$}}
  & \ns{\cell{5.02}{5.40}{0.38}{77.0}{$p\!=\!.104$}}
  & \sig{\cell{4.68}{5.23}{0.55}{81.0}{$\mathbf{p\!=\!.046^{*}}$}}
  & \ns{\cell{4.98}{5.38}{0.40}{32.5}{$p\!=\!.115$}}
  & \ns{\cell{4.90}{4.95}{0.05}{105.5}{$p\!=\!.487$}}
  & \sig{\cell{4.87}{5.42}{0.55}{50.5}{$\mathbf{p\!=\!.013^{*}}$}}
  & \sig{\cell{5.04}{5.43}{0.39}{6223.5}{$\mathbf{p\!<\!.001^{***}}$}}
  \\ \midrule
 
\textbf{Visual}
  & \ns{\cell{4.73}{5.48}{0.75}{91.5}{$p\!=\!.055$}}
  & \sig{\cell{4.67}{5.42}{0.75}{74.5}{$\mathbf{p\!=\!.030^{*}}$}}
  & \sig{\cell{5.08}{5.73}{0.65}{47.0}{$\mathbf{p\!=\!.029^{*}}$}}
  & \ns{\cell{5.07}{5.60}{0.53}{74.5}{$p\!=\!.090$}}
  & \ns{\cell{5.40}{5.55}{0.15}{114.0}{$p\!=\!.683$}}
  & \ns{\cell{5.07}{5.22}{0.15}{111.5}{$p\!=\!.625$}}
  & \ns{\cell{4.87}{5.15}{0.28}{82.5}{$p\!=\!.614$}}
  & \ns{\cell{5.10}{5.45}{0.35}{39.5}{$p\!=\!.411$}}
  & \ns{\cell{5.00}{5.07}{0.07}{71.5}{$p\!=\!.812$}}
  & \ns{\cell{4.85}{5.30}{0.45}{79.5}{$p\!=\!.125$}}
  & \sig{\cell{4.98}{5.40}{0.42}{7485.5}{$\mathbf{p\!<\!.001^{***}}$}}
  \\ \midrule
 
\textbf{Overall}
  & \sig{\cell{4.90}{5.40}{0.50}{380.0}{$\mathbf{p\!=\!.032^{*}}$}}
  & \sig{\cell{4.94}{5.49}{0.55}{278.5}{$\mathbf{p\!=\!.007^{**}}$}}
  & \sig{\cell{5.23}{5.71}{0.48}{213.5}{$\mathbf{p\!=\!.013^{*}}$}}
  & \sig{\cell{5.22}{5.63}{0.41}{258.0}{$\mathbf{p\!=\!.024^{*}}$}}
  & \sig{\cell{5.13}{5.62}{0.48}{288.5}{$\mathbf{p\!=\!.025^{*}}$}}
  & \ns{\cell{5.04}{5.31}{0.27}{381.5}{$p\!=\!.182$}}
  & \ns{\cell{4.78}{5.19}{0.42}{331.0}{$p\!=\!.084$}}
  & \ns{\cell{5.04}{5.41}{0.38}{138.0}{$p\!=\!.083$}}
  & \ns{\cell{4.95}{5.01}{0.06}{350.0}{$p\!=\!.573$}}
  & \sig{\cell{4.86}{5.36}{0.50}{259.5}{$\mathbf{p\!=\!.006^{**}}$}}
  & \sig{\cell{5.01}{5.41}{0.40}{27760.0}{$\mathbf{p\!<\!.001^{***}}$}}
  \\ \bottomrule
\end{tabular}
}
\end{table*}
}
 
\newcommand{\tableLayer}{%
\begin{table*}[ht]
\caption{%
  Wilcoxon signed-rank tests comparing \system (VT) vs.\ baseline (B) ratings
  (1--7 scale), stratified by layer number and question.
  Each cell reports the paired means and test result:
  B\,=\,baseline mean, VT\,=\,\system mean, $\Delta$\,=\,VT\,$-$\,B,
  $W$\,=\,Wilcoxon statistic, $p$\,=\,two-tailed $p$-value.
  The \emph{Overall} column pools all questions within each row.
  Layer~1 does not include Correct Scope as it does not have a parent layer.
  Significant results ($p < .05$) are \textbf{bolded}.
  $^{*}p < .05$;\;$^{**}p < .01$;\;$^{***}p < .001$.
}
\Description{Comparison of VideoTrail and baseline ratings by hierarchy layer. The table reports paired mean ratings and Wilcoxon signed-rank test results for each evaluation criterion at layers one, two, and three.}
\label{tab:vt_vs_baseline_layer}
\sffamily
\renewcommand{\arraystretch}{1.3}
\setlength{\tabcolsep}{4pt}
\scriptsize
\resizebox{\textwidth}{!}{%
\begin{tabular}{p{0.05\textwidth}
                p{0.07\textwidth} p{0.07\textwidth} p{0.07\textwidth}
                p{0.07\textwidth} p{0.07\textwidth} p{0.07\textwidth}
                p{0.07\textwidth} p{0.07\textwidth} p{0.07\textwidth}
                p{0.07\textwidth} p{0.07\textwidth}}
\toprule
\thead{Layer} &
  \thead{Visual \\ Coverage} &
  \thead{Dialogue \\ Coverage} &
  \thead{Faithfulness} &
  \thead{Coherence} &
  \thead{Segment \\ Diversity} &
  \thead{Logical \\ Boundaries} &
  \thead{Correct \\ Granularity} &
  \thead{Correct \\ Scope} &
  \thead{Image \\ Utility} &
  \thead{Text \\ Utility} &
  \thead{Overall} \\
\midrule
 
\textbf{Layer 1}
  & \ns{\cell{5.15}{5.25}{0.10}{42.0}{$p\!=\!.799$}}
  & \ns{\cell{5.10}{5.40}{0.30}{31.5}{$p\!=\!.314$}}
  & \ns{\cell{5.75}{5.85}{0.10}{26.0}{$p\!=\!.507$}}
  & \ns{\cell{5.75}{5.80}{0.05}{25.0}{$p\!=\!.782$}}
  & \ns{\cell{5.75}{6.00}{0.25}{39.0}{$p\!=\!.641$}}
  & \ns{\cell{5.75}{5.40}{-0.35}{25.0}{$p\!=\!.249$}}
  & \ns{\cell{5.30}{5.50}{0.20}{18.5}{$p\!=\!.632$}}
  & ---
  & \ns{\cell{5.20}{4.90}{-0.30}{42.0}{$p\!=\!.496$}}
  & \ns{\cell{5.25}{5.30}{0.05}{41.0}{$p\!=\!.740$}}
  & \ns{\cell{5.44}{5.49}{0.04}{2843.5}{$p\!=\!.750$}}
  \\ \midrule
 
\textbf{Layer 2}
  & \ns{\cell{5.50}{5.20}{-0.30}{61.5}{$p\!=\!.290$}}
  & \ns{\cell{5.45}{5.38}{-0.08}{47.5}{$p\!=\!.751$}}
  & \ns{\cell{5.53}{5.48}{-0.05}{31.0}{$p\!=\!.857$}}
  & \ns{\cell{5.58}{5.43}{-0.15}{55.0}{$p\!=\!.771$}}
  & \ns{\cell{5.65}{5.45}{-0.20}{56.5}{$p\!=\!.546$}}
  & \ns{\cell{5.55}{5.40}{-0.15}{45.0}{$p\!=\!.633$}}
  & \ns{\cell{5.40}{5.10}{-0.30}{53.0}{$p\!=\!.257$}}
  & \ns{\cell{5.55}{5.28}{-0.28}{39.0}{$p\!=\!.227$}}
  & \ns{\cell{5.48}{5.23}{-0.25}{20.5}{$p\!=\!.472$}}
  & \ns{\cell{5.20}{5.38}{0.18}{45.5}{$p\!=\!.404$}}
  & \ns{\cell{5.49}{5.33}{-0.16}{4471.5}{$p\!=\!.102$}}
  \\ \midrule
 
\textbf{Layer 3}
  & \sig{\cell{4.05}{5.75}{1.70}{6.0}{$\mathbf{p\!<\!.001^{***}}$}}
  & \sig{\cell{4.28}{5.70}{1.43}{10.0}{$\mathbf{p\!=\!.001^{***}}$}}
  & \sig{\cell{4.43}{5.80}{1.38}{10.5}{$\mathbf{p\!=\!.002^{**}}$}}
  & \sig{\cell{4.33}{5.65}{1.33}{7.0}{$\mathbf{p\!=\!.002^{**}}$}}
  & \sig{\cell{4.00}{5.40}{1.40}{1.5}{$\mathbf{p\!=\!.001^{**}}$}}
  & \sig{\cell{3.83}{5.13}{1.30}{14.0}{$\mathbf{p\!=\!.002^{**}}$}}
  & \sig{\cell{3.63}{4.98}{1.35}{11.0}{$\mathbf{p\!=\!.002^{**}}$}}
  & \sig{\cell{4.53}{5.55}{1.03}{4.5}{$\mathbf{p\!=\!.002^{**}}$}}
  & \sig{\cell{4.18}{4.90}{0.73}{19.0}{$\mathbf{p\!=\!.019^{*}}$}}
  & \sig{\cell{4.13}{5.40}{1.28}{9.0}{$\mathbf{p\!=\!.002^{**}}$}}
  & \sig{\cell{4.14}{5.43}{1.29}{772.5}{$\mathbf{p\!<\!.001^{***}}$}}
  \\ \bottomrule        
\end{tabular}
}
\end{table*}
}

\begin{abstract}
People often watch videos to find information, but locating relevant moments can be difficult and time-consuming.
Existing strategies like scrubbing rely primarily on visual scanning and linear search, providing limited support for goal-directed navigation. 
We present \system, an AI-powered browser extension that supports information seeking in videos through a hierarchical visualization.
Inspired by information foraging theory, \system surfaces semantic and structural cues as information scent, creating an explorable roadmap of segments.
The system is multimodal and adaptive, prioritizing visual or textual scent based on the video's content.
In a user study ($N=16$), \system helped participants efficiently locate key moments and enabled new navigation strategies compared to traditional timelines, highlighting how exposing an adaptive video structure can enhance goal-directed viewing.
\end{abstract}


\figureTeaser

\received{20 February 2007}
\received[revised]{12 March 2009}
\received[accepted]{5 June 2009}

\maketitle

\section{Introduction}
People often watch online videos to seek information --- whether learning how to fix a toaster, planning a trip, or reviewing a math concept.
However, video platforms such as YouTube are primarily designed for linear playback and passive consumption~\cite{weir2015learner,liu2018concept,kim2014data}, providing little support for \textit{goal-directed viewing}, where users watch with a specific objective rather than pure entertainment~\cite{seo2021active,pavel2014video}.
As a result, much of this information remains locked inside black boxes of video.
Even with a clear viewing goal, finding relevant moments is time-consuming and challenging, often requiring manual scrubbing through long, unstructured timelines~\cite{houten2004video,yang2025video}.

In this paper, we explore how video interfaces can return control to users.
Our approach is guided by two complementary ideas.
First, recent advances in multimodal AI enable richer video understanding and new forms of interaction~\cite{tang2026video,li2025videochat,maaz2024video,hua2025v2xum}, which we leverage to support goal-directed viewing.
Second, to ensure that we design useful interactions for viewers, we draw on information foraging theory (IFT)~\cite{pirolli1999information,pirolli2005sensemaking}, which models information seeking as a selective process of evaluating cues or ``scent.''
Applied to video, this process is known as \textit{video foraging}~\cite{houten2009searching,houten2004video},
and \edit{we extend this framing to a complementary task we call \textit{reforaging}: rediscovering information in familiar videos.}

We conducted a formative study ($N=6$) to examine current foraging practices \edit{and collect feedback on our initial design ideas.}
Consistent with IFT, users rarely watched videos sequentially; instead, they iteratively skimmed, jumped, and refined their search to locate relevant moments. 
However, existing skimming strategies often rely on a single modality (\eg visual scrubbing)~\cite{pavel2014video,matejka2013swifter,fraser2020temporal}, which provides weak information scent for certain videos (\eg interviews).
They also largely assume linear navigation~\cite{jin2017elastic,christel1999adjustable,truong2021automatic}, while goal-directed viewing is selective and uneven
--- requiring deep engagement with some segments while skipping others entirely. 
Consequently, users may struggle to efficiently and accurately assess the informational 
value of clips \edit{in both familiar and unfamiliar videos}~\cite{christel2008supporting,albertson2013thumbnails}.

To address these challenges, we introduce \system, an AI-powered browser extension for foraging and reforaging through \textit{semantic navigation}.
Instead of a linear timeline, \system visualizes a video as a hierarchy of segments (\autoref{fig:teaser}), creating an explorable and searchable roadmap.
The hierarchical representation enables users to maintain a high-level overview while progressively drilling down into segments of interest, supporting diverse viewing goals.
To guide foraging, \system adaptively weights visual vs. textual information scent based on video content.
It also generates contextual collections and color scales to further aid exploration.
Underlying the system is a multimodal LLM (MLLM) pipeline that automatically extracts detailed semantic structure from videos.

We evaluated \system through a two-part user study ($N=16$) combining controlled comparisons with YouTube and open-ended exploration of participant-selected videos.
Results show that \system's adaptive, hierarchical summaries support foraging across various video types, lowering the cognitive cost of exploration and allowing users to quickly locate new information and revisit previously encountered content.
Participants also developed a clearer understanding of video structure and discovered unexpected moments of interest, suggesting further opportunities for AI-powered visualizations to aid navigation and sensemaking of temporal data.

Our contributions include:
\begin{itemize}
    \item \textbf{A formative study} characterizing the limitations of current video interfaces and design opportunities for goal-directed viewing through the lens of IFT.
    \item \textbf{\system, a browser extension} that supports \edit{both} foraging and reforaging through multimodal, hierarchical visualizations that adapt to video content.
    \item \textbf{User study findings} demonstrating how surfacing semantic structure and contextual information scent enhances video skimming and navigation beyond linear timelines.
\end{itemize}

\section{Related Work}
\edit{As AI makes it easier and faster to extract structured information from videos~\cite{tang2026video,hua2025v2xum}, we focus on the complementary challenge of making that structure useful for goal-directed viewing.
Our approach bridges work in video summarization and representation, structured video navigation, AI-assisted video understanding, and information foraging.}

\subsection{Video Summarization and Visualization}
Video summarization approaches are typically categorized as extractive or abstractive~\cite{alaa2024video,christel1999adjustable,gupta2023comprehensive}.
\textit{Extractive} methods select and preserve content from the original video, often through visual formats such as image collages~\cite{chen2012visual,mei2009video,goldman2006schematic}, storyboards~\cite{furini2007visto,macer1996video,narasimhan2022tl,hurst2012history,dawes2004cinema}, or highlight reels~\cite{vo2025integrate,jin2017elastic,lin2024videogenic,kim2025generating,narasimhan2021clip,patel2024your,wu2022intentvizor,wang2024podreels}.
Conversely, \textit{abstractive} approaches generate new representations, such as stylized comic strips~\cite{pesko2019comixify,wang2012movie,boreczky2000interactive,calic2007efficient,wang2025video} or text summaries~\cite{wang2022record,palaskar2019multimodal,dilawari2019asovs}.

Recent work pairs extractive keyframes with abstractive captions for richer, multimodal context~\cite{barua2025lotus,hua2025v2xum,lee2025video,kudo2023challenging}.
Visual analytics offers another approach for summarizing videos~\cite{jorge2013exploring,zeng2020emotioncues}, for example, through shot clusterings~\cite{junyent2015video}, character networks~\cite{vicol2018moviegraphs,zhang2014videograph}, and annotated timelines~\cite{kurzhals2016visual,li2024pycinemetrics}.
\edit{Building on this work, we design interactive, multimodal video representations to surface information scent and support active, goal-directed exploration.} 

\subsection{Navigating Videos Through Structured Representations}
Prior interfaces augment navigation by enhancing video timelines with grid-based scrubbing~\cite{matejka2013swifter}, dynamic speed adjustments~\cite{pongnumkul2010content}, or interaction overlays~\cite{kim2014crowd,alhajri2014visualization}.
Video Tapestry~\cite{barnes2010video}, SpiralTape~\cite{liu2016interactive}, and VideoTrees~\cite{jansen2008videotrees} use zoomable, multi-scale representations to navigate videos, but rely primarily on visual cues.

Other systems expose higher-level semantic structure to facilitate skimming and navigation, often via video chapters~\cite{Ventura_2025_CVPR,fraser2020temporal,pu2025arc} (\eg YouTube~\cite{youtubechapters}) or hierarchical summaries~\cite{girgensohn2011adaptive,jiang2016hierarchical,van2024making,weir2015learner}.
Structured interfaces such as Video Digests~\cite{pavel2014video} and Visual Transcripts~\cite{shin2015visual} support lecture videos, while others focus on music~\cite{wang2021soloist} or makeup tutorials~\cite{truong2021automatic}.
Recent works also explore nonlinear concept maps~\cite{liu2018concept} and dependency graphs~\cite{chen2024tuto,yang2022improving}.
However, these systems often rely on domain-specific heuristics or use fixed (\eg two-level) hierarchies.
\edit{With \system, we investigate how to surface multimodal video structure that adapts to different content types and enables navigation at multiple granularities based on user goals.}

\subsection{AI-Assisted Video Understanding}
Multimodal AI now enables on-demand video summaries without manual annotation~\cite{weir2015learner,pavel2014video,kim2014crowd,psallidas2023video} or specialized pipelines~\cite{truong2021automatic,chen2012visual}.
Recent MLLM-based systems support conversational interaction, allowing users to query videos and locate specific timestamps~\cite{li2025videochat,maaz2024video,tang2026video}.
WatchWithMe~\cite{chi2025watch} facilitates guided watching of review videos, StoryNavi~\cite{xu2024storynavi} enables narrative-based reconstruction, and TLDW~\cite{zhang2025why} generates personalized highlight reels.
YouTube also recently integrated an AI-powered ``ask'' feature~\cite{shrivastava2025youtube}. 

While these systems prioritize quick, text-based access to content, complementary work uses AI to restructure video into interactive representations that enable deeper comprehension and reasoning~\cite{xia2024video}.
For instance, VideoMix~\cite{yang2025video} and TutoAI~\cite{chen2024tuto} extract video structure to support cross-video summarization and mixed-media tutorial authoring.
\edit{Our work instead uses AI-extracted structure to support ongoing information seeking within a single video, helping viewers assess segment relevance and move between levels of abstraction when deciding where to explore next.}

\subsection{\edit{Information Foraging and Reforaging}}
Information foraging theory (IFT) models how people seek information on the web by following ``scent'' --- cues that signal value --- across ``patches,'' or structural units of information~\cite{pirolli1999information,pirolli2005sensemaking}.
IFT has informed the design of systems for online research~\cite{rachatasumrit2021for}, document exploration~\cite{fok2024marco}, and LLM-assisted sensemaking~\cite{suh2023sense}.
Extending IFT to video,~\citet{houten2009searching} introduced \textit{video foraging}, which frames contiguous video segments as patches.
Because video is time-based, navigation is particularly costly, making strong information scent critical~\cite{houten2004video}.
While foraging occurs both within and across videos~\cite{houten2009searching}, our focus is the former.

\edit{Much IFT work focuses on finding information in new environments~\cite{rachatasumrit2021for,suh2023sense,fok2024marco,houten2004video}, but seeking previously encountered content often relies on different cues and strategies~\cite{tyler2010large,capra2005using}. 
Video interfaces have supported revisiting information through personal viewing histories~\cite{alhajri2014visualization}, bookmarks~\cite{kim2014data}, and synchronized views of scripts and summaries~\cite{pavel2015scene}.
We extend this work by foregrounding the task of \textit{reforaging}, \ie refinding moments in familiar videos.
Specifically, we explore how generative AI can support both foraging and reforaging by exposing semantic structure and adaptive multimodal cues as information scent.}

\section{\edit{Formative Study: Exploring Design Opportunities for Video Foraging and Reforaging}}\label{sec:formative}
\edit{We interviewed six participants (\textbf{F1-6}) at~\location to (1) understand existing workflows and challenges when seeking information within videos and (2) collect feedback on an early prototype, which we used as a probe~\cite{boehner2007hci,graham2008probes} to evaluate our initial design ideas.}
In each 30-minute session, participants first described their current practices for goal-directed viewing on YouTube, then explored two self-selected videos with our prototype: one familiar and one unfamiliar.
Videos ranged from visual how-tos (\eg recipes, art tutorials) and math explainers to dialogue-driven content (\eg video essays, interviews), allowing us to examine how a hierarchical, multimodal visualization could support diverse content types and information needs.

\subsection{\edit{Video Foraging Practices \& Challenges}}

Users distinguished goal-based consumption from entertainment-driven viewing, which is typically sequential:
\myquote{Comedy shows [and] movies always have surprising turning points... so I watch them from start to finish, because that's the art they're trying to make} (F2).
In contrast, during \textbf{information seeking}, participants rarely watched videos linearly. 
Instead, they selectively navigated and sampled segments to identify relevant or interesting content, especially in longer videos ($n=3$).
This behavior mirrors the process of iteratively assessing information scent to decide how to navigate between patches (\ie video segments) in IFT~\cite{pirolli2005sensemaking,houten2009searching}.

All participants described \textbf{skimming} as their main way of finding information in videos, using strategies such as fast-forwarding or increasing playback speed ($n=3$), scrubbing the timeline ($n=2$), or reading the transcript ($n=1$).
However, three recurring challenges emerged from these approaches.

\subsubsection{Weak information scent from single-modality cues}
Five participants described skimming as a ``blind'' trial-and-error process because contextual cues are typically \textbf{limited to a single modality} --- \eg visual thumbnails for scrubbing or text for transcripts.
As videos are inherently multimodal, this obscures key signals about a segment's informational value~\cite{jorge2013exploring,wingstedt2010narrative}. 
F3 shared that \myquote{lectures in particular} are challenging to skim, \myquote{because just scrubbing through the video, it doesn't look that different.} 
Similarly, F4 wanted more semantic context when navigating video essays: \myquote{I'd like to see at a glance, what the overall argument is. I don't want to constantly click to get that information.}

\subsubsection{Limited support for nonlinear information seeking}
While current skimming methods enable faster traversal of time, they largely \textbf{assume linear, uniform consumption} and do not account for differences in informational value across segments ($n=4$).
This creates a mismatch with goal-directed viewing, which participants described as inherently selective and nonlinear.
Rather than traversing chronologically, they wanted to filter segments by intent: \myquote{I'm usually looking for something pretty specific so it'd be great if I could just explore those aspects [of the video]} (F6). 
However, with existing interfaces, users frequently had to probe multiple clips before finding relevant content, increasing interaction cost.
As F3 explained, \myquote{it takes me forever to find the interesting parts... there's just so much random stuff to sort through.}

\subsubsection{\edit{Challenges persist when reforaging familiar videos.}}
\edit{Prior familiarity did not eliminate these navigation challenges.
For example, F1 described how he \myquote{often like[s] to revisit} math explainers, but still has \myquote{to guess where things are} each time.
Participants also emphasized the \textbf{need for adaptive depth}, especially when rewatching content.
While scrubbing a familiar video, F5 noted, \myquote{Actually, maybe I'd skip this part since I already know it from [before], but here, I'd want to dive deeper to learn more.}
Thus, reforaging involves not only relocating previously encountered information, but also deciding what to skip, revisit, or explore further.
}

\subsection{\edit{Design Probe Feedback}}
These findings suggest opportunities to better support \textit{nonlinear, selective} information seeking, \edit{whether viewers are exploring a video for the first time or returning to familiar content.
To explore this space, we used our early prototype to probe two initial design ideas: representing videos as a navigable semantic hierarchy, and combining visual and textual cues to strengthen information scent (see~\autoref{app:form_study}).} 

\subsubsection{\edit{Strengths}}
All six participants found the \textbf{hierarchical visualization} useful --- particularly for foraging long-form content.
F2 felt it offered \myquote{more precise [and] convenient} navigation because of the \myquote{semantic segments that reveal the video's structure.}
F6 appreciated being able to focus on specific portions of an art video: \myquote{I like how I can just expand the section I'm interested in, and then maybe I could take a screenshot and use that as my mini tutorial... [that] would be super useful for revisiting in the future too.}
Participants also valued the system's diverse, \textbf{multimodal cues} ($n=5$). 
F1 said, \myquote{Even just seeing the keywords and descriptions [along with] the keyframe was useful for finding specific information.}

\subsubsection{\edit{Areas for improvement}}
\edit{The probe also revealed opportunities to refine our design.}
Four participants asked to \textbf{customize scent:} \myquote{For some videos, I don't think the keywords would be that helpful. But for something like this [video essay], I'd actually want to see even more text} (F5).
\edit{Participants also suggested ways to reduce visual complexity, \eg prioritizing \textbf{progressive expansion} instead of showing the full video hierarchy at once ($n=4$).}
Additionally, three wanted to annotate and save segments, suggesting more iterative, personalized ways to seek and revisit information.
\edit{\autoref{app:form_study} includes a more detailed account of how our prototype evolved after the formative study.}

\subsection{Design Goals}\label{sec:goals}
\edit{Together, participants' existing practices, probe feedback, and IFT informed three design goals for supporting both video foraging and reforaging:}
\begin{enumerate}[start=1,label={\bfseries D\arabic*.}] 
    \item Help viewers locate and relocate relevant segments.
    \item Enable nonlinear, semantic navigation with progressive disclosure.
    \item Surface salient cues for both dialogue- and visual-driven content.
\end{enumerate}

To address these goals, we developed a data pipeline (\autoref{sec:pipeline}) and \edit{final interface} (\autoref{sec:system}) that transform unstructured videos into navigable, hierarchical structures.
\edit{Our core contribution is the latter, but we first describe the pipeline as it defines the semantic representation our system visualizes and interacts with.}

\section{Data Extraction Pipeline: Surfacing Hierarchical Video Structure With AI}\label{sec:pipeline}
\figurePipeline
\edit{\system operates over a hierarchical video representation.
To generate this representation automatically, we use an MLLM-based pipeline that recursively segments a video and captures content at multiple levels of granularity.
}

\subsection{Pipeline Overview}
Given any YouTube video as input, our pipeline outputs multimodal, hierarchical segment data and a list of suggested collections (\autoref{fig:pipeline}). 
Data extraction is fully automated and consists of four key steps.
Thresholds (\eg max segments per layer) were tuned to balance coverage and skimmability, but can be adjusted.

\boldni{Step 1: Determine overall video characteristics.}
First, we extract high-level metadata, including the input video's primary \textit{modality} (visual or dialogue)\footnote{We currently adopt a binary notion of modality based on our formative findings.}, determined by which channel plays a more central role in conveying content.
We use this to prioritize words or visuals during analysis, and later guide how information is presented to users (\textbf{D3}).
The model then identifies 2-4 \textit{semantic dimensions} that describe how the video varies (\eg ``actionability'' in~\autoref{fig:teaser}).
Each dimension is defined by endpoints (\eg ``theoretical'' to ``highly actionable'') and mapped to a color scale, facilitating comparison along these axes (\textbf{D1}).
To support skimming, we generate a 1-2 sentence \textit{summary} and 3-5 \textit{keywords}.
Keywords cover aspects such as \textcolor{visual}{visual}, \textcolor{dialogue}{dialogue}, \textcolor{sound}{sound}, \textcolor{mood}{mood}, or \textcolor{theme}{theme}~\cite{jorge2013exploring,wingstedt2010narrative}, and each comes with a short description of how it relates to the video.

\boldni{Step 2: Segment first layer.}
The video is partitioned into timestamped segments to create an initial semantic outline. 
Each segment returns multimodal data: a \textit{title}, \textit{keyframe}, \textit{summary}, 1-3 \textit{keywords}, and \textit{ratings} for the specified dimensions (range: [0,1] or [-1,1] for diverging traits like sentiment). 
Keyframes are selected based on relevance to the title and summary.
To improve skimmability, the LLM crops busy keyframes (\eg lots of text or complex elements) to focus on salient regions.
We validate segments by merging them if too many are returned ($>$7) and adjusting highly uneven boundaries (\ie one segment is $>$2.5x longer than the others).
Finally, we select 2-3 \textit{highlights} to enable quick access to key moments (\textbf{D1}). 

\boldni{Step 3: Split segments recursively to form deeper layers.}
We repeat this process to split segments into subsegments, forming a multi-level hierarchy (\textbf{D2}). 
Splitting is context-aware: subsegments inherit their parent's context while adding finer detail.
Each parent segment can have up to 5 children and 1-2 highlights.
To avoid over-fragmentation, we stop when the max depth is reached (4 layers), segments become too short ($<$10 seconds per YouTube chapter guidelines~\cite{youtubechapters}) or the model determines that further splitting does not add meaningful structure (\eg intro or end screen)~\cite{Ventura_2025_CVPR,pu2025arc}.

\boldni{Step 4: Generate suggested collections.}
Using the resulting hierarchy, we generate 3-5 \textit{suggested collections}; each groups semantically related clips that represent a cross-cutting theme (\eg ``Culinary Tour \& Highlights'' in a travel vlog).
Collection segments do not need to be contiguous, offering another nonlinear way to explore the video (\textbf{D1}).
The final video data is outputted as JSON.

\subsection{Implementation \& Performance}\label{sec:pipeline_imp}
The pipeline, implemented in Python, uses \texttt{gemini-3-flash-previ \allowdisplaybreaks ew}~\cite{geminiflash} to extract most video data, passing YouTube URLs directly as input.
We extract keyframes with yt-dlp~\cite{ytdlp} and use \texttt{claude-haiku \allowdisplaybreaks -4-5}~\cite{claudehaiku} to identify salient cropping regions. 
Most pipeline steps are parallelized to reduce latency (\eg segment splitting, keyframe extraction, collection generation), and prompts are included as Supplementary Materials. 

With this pipeline, a 30-minute video (\eg \href{https://www.youtube.com/watch?v=9Zz2KrBDXUo}{\textit{DWS News} interview}) takes $\sim$7 minutes to process and generates $\sim$\textbf200 segments across four layers; a 3-hour video (\eg \href{https://www.youtube.com/watch?v=YQGai2PVHBs}{\textit{Insider Food} documentary}) takes $\sim$15.5 minutes and yields $\sim$600 segments.
\edit{While this latency limits lightweight, on-demand use, our focus is prototyping useful downstream interactions over the resulting structure (assessed in \autoref{sec:results}) rather than optimizing preprocessing speed.
We also expect data generation to become faster as multimodal inference improves.
For reference, we evaluated our pipeline against a recent baseline for video chapter and summary generation and found comparable performance; full methods and results are reported in~\autoref{app:tech_eval}.}

\figureInterface
\section{\system Design}\label{sec:system}
\system is an AI-powered browser extension that supports video foraging through enacting our design goals (\autoref{fig:interface}).
Leveraging structured outputs from our pipeline, \system decomposes videos into semantic segments and connects them into a hierarchical visualization.
Users navigate the hierarchy by following the information scent of segments, which adapts based on video content.
While individual layers may resemble YouTube chapters~\cite{youtubechapters}, \system extends this paradigm by making segments interactive and enabling exploration across multiple levels of detail~\cite{min2025malleable,van2024making,li2021hierarchical}.

\subsection{Interface Overview}
\system integrates three main views.
The left \textbf{YouTube Panel} preserves the native video player, description, and comments (\autoref{fig:interface}A).

On the right, we augment YouTube's UI with a custom interface.
At the top, we display the video's \textbf{Summary \& Collections}.
The \textit{summary} tab shows the global summary and keywords (\autoref{fig:interface}B), while 
\textit{collections} offers a place to save and organize segments of interest --- similar to bookmarking locations on a map~\cite{applemapsguides,googlemapslists}.

Below, the \textbf{Video Map} displays our main hierarchical visualization, which the user can expand, search, or customize (\autoref{fig:interface}C).
As the video plays, the active path in the hierarchy is highlighted. 
Conversely, selecting a segment in the Video Map updates playback to the corresponding moment, enabling rapid, context-aware navigation.

\subsection{Segment Nodes}
The base unit of our visualization is the \textbf{segment node}, which represents a contiguous segment (patch) in the video.
Each node consists of a \textit{title}, \textit{colorbar}, \textit{keyframe}, \textit{summary}, and segment-specific \textit{keywords} that together shape its scent (\textbf{D1}).
The colorbar indicates the segment's rating for the current dimension (\eg ``activity pace'' in~\autoref{fig:interface}C).
Hovering over the keyframe plays a 5-second preview, and segments can be saved via the ``collect'' button.

\figureAdaptive
\subsubsection{Adaptive scent}
Segment nodes surface or hide information to adapt to video content and preserve skimmability (\textbf{D3}). 
For \textit{visual-driven} videos (\eg travel vlog in \autoref{fig:adaptive}), \system emphasizes the keyframe (visual cue) and hides summary text, using keywords to provide quick context.
For \textit{dialogue-driven} content (\eg talking head video), we prioritize the segment summary (text cue), while only showing a partial keyframe, which likely carries less scent.
Keywords are also hidden in this case to reduce visual clutter. 

Nodes support \textit{semantic zoom}~\cite{suh2024luminate,suh2023sense}, where information density adjusts based on zoom level (\textbf{D2}).
As the user zooms in, all segments (regardless of video type) progressively reveal their full keyframe and summary text (\autoref{fig:zoom}).

\subsection{From Segments to Hierarchy}
Segments are organized into \textbf{layers}, which are connected to form the video \textbf{hierarchy} (\textbf{D2}).
The top layer provides a high-level overview, while deeper layers expand nodes into subsegments, revealing increasingly fine-grained detail (\autoref{fig:teaser}B).
Users can click any node marked ``expand'' to navigate; this reveals its subsegments in the next layer.
Clicking again collapses nodes up one level.
Expansion dynamically adjusts the visualization layout to preserve segment chronology (\autoref{fig:interface}C).
Edges along the active path are highlighted with a darker opacity to preserve context.
To focus exploration and prevent visual overload, only one node per layer can be expanded at a time.
\edit{Our formative probe explored a ``full view'' exposing the entire hierarchy simultaneously, but participants found it difficult to navigate for more complex videos (see \autoref{app:form_study} for more details).}

\figureSearch
\subsection{Exploring the Hierarchy}
\system provides support for both targeted information-finding and open-ended discovery (\textbf{D1}).

\subsubsection{Semantic search}
Users can locate specific information via \textit{semantic search} (\autoref{fig:interface}C).
Expanding the search box allows inputting a natural language query or viewing question suggestions (\autoref{fig:search}A).
Rather than returning an isolated textual answer, \system identifies relevant segments (\autoref{fig:search}B) and spotlights them in the hierarchy (\autoref{fig:search}C).
This enables users to interpret results in context using the video's structure, explore nearby segments, or jump directly to the corresponding moment in the video (\textbf{D2}).
Users can navigate between results via the search box, which expands the path to the corresponding segment, or by traversing the visualization itself.

\subsubsection{\edit{Additional exploration controls}}
At the top of the Video Map, users can switch the attribute encoded by colorbars via the \textit{visualize} dropdown,
or add custom scales (via the ``+'' button) to specify new dimensions to explore~\cite{yeh2026story} (\autoref{fig:interface}C).
Users can also \textit{sort} nodes based on their semantic ratings, surfacing relevant moments (\eg most energetic segments).

For less targeted goals (\eg ``I want to discover interesting segments''), the \textit{highlights} toggle reveals recommended segments at each layer of the hierarchy, forming a guided exploration path (\textbf{D2}, \autoref{fig:highlights}).
Additionally, the \textit{collections} view displays system-suggested collections that cover key video themes (\autoref{fig:collections}). 
Upon opening a collection, corresponding segments in the Video Map are highlighted --- similar to search.

\subsection{Extension Implementation}\label{sec:implementation}
\system is a browser extension with a TypeScript/React frontend and Python/Flask backend.
Our visualization is built on top of React Flow~\cite{reactflow}, with custom Sankey-style SVG edges.
Video data (from our pipeline) is served from S3.
We use \texttt{gemini-3-flash-pre \allowdisplaybreaks view}~\cite{geminiflash} to power all on-the-fly AI features, \ie search and custom colors. 
Both functions take the hierarchical structure as input, along with the search query or new rating dimension.
For search, we perform a heuristic, breadth-first traversal.
At each level, we evaluate relevance and only expand relevant nodes, pruning the remaining subtrees.
Prompts are included as Supplementary Materials.

\section{User Study}\label{sec:study}
To evaluate \system's support for video foraging \edit{and reforaging}, we conducted a two-part user study consisting of (1) a controlled comparison with YouTube and (2) an open-ended exploration task using participant-selected videos.
Our study was guided by three research questions:

\begin{enumerate}[start=1,label={\bfseries RQ\arabic*.}] 
    \item Does \system effectively support information seeking inside videos, and how does it compare to existing \edit{workflows}?
    \item Which aspects of \system do users find most useful for foraging and reforaging videos?
    \item How do users perceive the AI-generated video structure? 
\end{enumerate}

\subsection{Study Design}
The study consisted of two complementary phases.
During the \textbf{comparison task}, participants completed standardized information finding tasks using \system and YouTube, allowing us to compare both workflows under controlled conditions.
\edit{For an ecologically valid, holistic comparison, we selected YouTube as a mature, widely used platform that all participants already used for goal-directed viewing; \system also directly augments its interface.
Participants could freely use any available YouTube feature --- including the AI chat~\cite{shrivastava2025youtube} --- to reflect realistic use.}

For the \textbf{exploration task}, participants selected their own videos and goals to explore with \system, capturing interaction strategies across varied content in a more naturalistic setting.

\subsection{Participants}
We recruited 16 participants (\textbf{P1-16}), aged 18-30, from mailing lists at~\location. 
All participants watched YouTube at least once a week, with 12 watching on a daily basis.
Their strategies for finding information in videos included scrubbing the timeline ($n=8$), jumping to timestamps in the description or comments ($n=6$), adjusting playback speed or fast forwarding ($n=5$), or simply watching the full video ($n=5$). 
Four participants also leveraged YouTube's ``most replayed'' visualization to find interesting moments.
None had previously used LLMs to navigate or understand video content. 

\subsection{Materials}
Our study focuses on long-form videos based on formative feedback indicating this as an ideal use case (\autoref{sec:formative}), and to ensure that participants could not just complete tasks by watching a video in full. 
All videos were preprocessed and loaded into \system prior to the study session.
The full list of videos and questions used across both tasks is provided in the Supplementary Materials.

\subsubsection{Comparison task.}
We wanted to evaluate utility for videos with different primary modalities.
Thus, our corpus included two \textit{visual}-driven videos (travel vlog, recipe) and two \textit{dialogue}-driven videos (product review, advice) with similar durations ($\sim$35 minutes) and information complexity.
The advice video and travel vlog are shown in \autoref{fig:teaser} and \autoref{fig:interface}, respectively.
We included two stimuli per category to reduce confounding effects due to a particular video's characteristics (\eg one being inherently easier to find information in).
Following~\citet{yang2025video}, each participant explored one video of each category during the study.
We also ensured that participants had not previously watched their assigned videos.

For each video, participants completed four information-seeking tasks of increasing complexity:
\begin{enumerate}[label=\arabic*.]
    \item \textbf{Overview:} provide a 1-2 sentence summary of the video.
    \item \textbf{Retrieval:} locate a specific segment. 
    \item \textbf{Evidence:} explain a claim made in the video. 
    \item \textbf{Synthesis:} compare two concepts.
\end{enumerate}
These categories were informed by prior evaluations of video understanding and navigation systems~\cite{shin2015visual,truong2021automatic,chi2025watch}.
A fixed set of questions was curated for each video through iterative testing to ensure answers were discoverable within a reasonable time frame.

\subsubsection{Exploration task.}
Participants chose two long-form videos ($\geq$20 minutes) to explore: one familiar and one unfamiliar.
Selected videos averaged 41.2 minutes in duration (\textit{min:} 20.5 minutes, \textit{max:} 3.2 hours), ranging from a cheese documentary and politician interview to a luxury airline review and 3D modeling software tutorial.
The most common video types included explainers about math, history, or psychology ($n=9$), video essays about current events or popular media ($n=5$), reviews about different products or experiences ($n=5$), and how-to tutorials ($n=4$).

\subsection{Procedure}
Sessions lasted one hour and took place over Zoom. 
We asked participants to think aloud throughout the study, and with their consent, we collected screen/audio recordings along with system logs for further analysis. 
Participants received \$20 in compensation.

\subsubsection{Pre-task interview (5 min)}
We first probed participants' strategies for finding information in YouTube videos and prior experience foraging videos with LLMs (reported above).

\subsubsection{Comparison task (25 min)}
Participants used both \system and YouTube to answer questions about two videos (one per interface).
Interface order and video assignment were counterbalanced.
Before using \system, participants received a five-minute tutorial.
After each condition, they filled out a brief Likert-scale survey probing utility and ease of use. 
Items were adapted from the System Usability Scale (SUS)~\cite{brooke1996sus} and NASA-TLX~\cite{hart1988development}, along with validated criteria for evaluating video interfaces~\cite{gupta2023comprehensive,pu2011user,airaldi2025measuring}.
Each condition took $\sim$10 minutes to complete (including the survey).

\subsubsection{Exploration task (20 min)}
Next, participants used \system to freely explore their two videos --- each toward a self-selected information goal ($\sim$10 minutes each).
The short time frame simulates realistic constraints when users may need to quickly glean information from long videos.

\subsubsection{Post-task interview (10 min)}
To conclude, we conducted semi-structured interviews with participants to collect qualitative feedback about using \system, including which features they found most useful.
We also asked for ideas to improve the system, and how participants perceived the AI-generated video structure.

\subsection{Data Analysis}
We adopted a mixed-methods approach to data analysis. 
For the comparison task, we measured performance by analyzing response accuracies and completion times across conditions.
Responses were scored against reference answers using a predefined rubric (see Supplementary Materials).
From screen recordings and system logs, we collected a quantitative account of feature usage for both tasks (\eg number of segments expanded, searches submitted). 
For all statistical comparisons, we used linear mixed models and exploratory simple effects analyses with BH correction. 
We also synthesized qualitative themes from think-aloud and interview transcripts.

\section{Results}\label{sec:results}
Overall, participants found \system useful for both exploratory and targeted information seeking. They described the system as \myquote{really clear and intuitive to use} for video foraging (P6) and \myquote{super powerful... it makes your life so much easier} (P11).
Some even called our visualizations \myquote{inspiring} (P12) and enjoyed how \system offered \myquote{a different way of looking at videos} (P3).
All 16 expressed interest in using the system again: \myquote{Is this extension available? I want it!} (P10) and \myquote{I'll miss \system when using YouTube} (P1).

\subsection{Does \system Effectively Support Information Seeking Inside Videos, and How Does It Compare to Existing Workflows? (RQ1)}\label{sec:rq1}
\edit{During the comparison study, we noticed distinct information-seeking workflows across the two interfaces. 
With YouTube, all participants used scrubbing, supplemented by playback-speed adjustments, timestamps or chapters, and transcript search, often switching approaches depending on the task.  
Three also tried the AI chat for the first time.
In contrast, foraging strategies with \system centered on the exposed video structure: 13 participants combined semantic search and hierarchical navigation, while three relied solely on the latter.
Across these workflows, \system showed consistent benefits for information finding, reflected in both participant perceptions and task performance.
} 
\subsubsection{Participants perceived \system as more useful for information seeking}
\system was rated significantly higher than YouTube across all measures ($p<.001$; \autoref{fig:likert}).\footnote{Full statistics are reported in \autoref{app:comp_results}.}
For \textit{dialogue-driven} videos, participants found it easier to obtain an overview ($p < .05$) and were more confident in finding information ($p < .01$) with \system.
For \textit{visual-driven} videos, they felt \system improved ease of information finding ($p < .01$), and was more helpful for navigation ($p < .05$) and understanding ($p < .01$).
Across both video types, \system better supported discovery of interesting moments 
(dialogue: $p <.05$, visual: $p < .01$) and was perceived as more enjoyable to use (dialogue: $p <.05$, visual: $p < .01$).

\figureLikert

\subsubsection{\system improved accuracy and reduced completion time} 
\system led to significantly \textbf{higher overall accuracy} ($p<.001$) and \textbf{faster total completion time} ($p<.05$), even with YouTube's AI chat feature (\autoref{fig:acc_times}).
Although chat responses were quick, users still needed to formulate questions, read the AI's answer, and verify the provided timestamps.
In contrast, \system allowed direct access to relevant segments through its hierarchical structure.
Within \system, node expansion was also often more efficient than searching with AI.
For example, while waiting for semantic search to run, P12 began exploring the Video Map, and found the answer before the search returned: \myquote{That's funny, it's actually just where I was. That's crazy I found it first.}

Participants answered \textit{evidence} questions significantly faster with \system ($p<.05$), which typically involved analyzing video dialogue that is difficult to skim via scrubbing.
The only case where \system was slower on average was for \textit{synthesis} questions in dialogue-heavy videos; these required integrating information from two disparate segments.
This may be due to the constraint of expanding one subtree at a time, though the difference was not significant.
Notably, \system still yielded higher accuracy for synthesis questions ($p<.05$).
Video type (dialogue or visual) did not significantly affect performance.

\subsection{Which Aspects of \system Do Users Find Most Useful For Video Foraging and Reforaging? (RQ2)}\label{sec:rq2}
To better understand how \system supports foraging, we analyzed interactions during open-ended exploration.

\figureAccTime
\edit{Participants adopted varied paths through the hierarchy:
nine broadly explored different segments across the Video Map, while five followed deeper, focused paths, and two tried a mix of both. 
We also observed hybrid strategies that combined \system's features with YouTube's, \eg expanding the hierarchy to a segment and scrubbing within that range ($n=2$), or using YouTube's description and timestamps to guide exploration in the Video Map ($n=3$), suggesting complementary workflows with existing tools.}

\edit{Node expansion remained the dominant foraging approach (users expanded a total of 433 nodes across 92 unique subtrees), followed by semantic search (46 queries).
11 participants browsed or created a collection, 10 viewed system highlights, and 13 explored semantic dimensions.}

\subsubsection{\edit{Three interactions were central to effective video foraging}}
\edit{Participants found \system's hierarchical navigation, structure-grounded search, and multimodal information scent particularly valuable during foraging, reflecting favorably on our design goals.}

\boldni{Hierarchical navigation supported selective exploration.}
13 participants valued how our visualization provided a \myquote{semantic map that shows you what's coming so I can easily find the most interesting parts [or] skip ahead} (P12, \textbf{D1}).
They appreciated the flexible navigation \myquote{between high and low level themes} (P15, \textbf{D2}), which offered more useful structure than \myquote{a timeline [that] can't show such granular categories} (P3).
In fact, six almost exclusively used the hierarchy to forage: \myquote{I feel like I almost didn't need the other features... the tree itself was so helpful [and] searching through it yourself is really fun} (P10).

\boldni{Grounding search in structure enhanced retrieval and discovery.}
Nine participants described \system's semantic search feature as a \myquote{game changer} (P4)
that \myquote{accurately found the right part even with just a loose phrase} (P15, \textbf{D1}).
Crucially, they appreciated how results were grounded in structure (\textbf{D2}): \myquote{It's cool how [the system] points to places to look through like a mini summary} (P9) and lets you \myquote{directly visualize and jump to it in the map} (P6).
Seven enjoyed how our search \myquote{pulls semantically similar segments [that] helps you discover other parts of the video you probably wouldn't have found yourself} (P13).

\boldni{Multimodal scent helped users assess relevance before watching.}
\system's rich, adaptive information scent (\textbf{D3}) enabled rapid skimming beyond visuals through
segment titles ($n=10$), summaries ($n=7$) and keywords ($n=4$).
P6 noted, \myquote{When I read these [text cues], I don't even need to click on the video.}
Semantic dimensions further supported personalized foraging ($n=6$, \textbf{D1}) by allowing \myquote{input to the system} to surface \myquote{contextual fluctuations within a video} (P14).
Users added dimensions to search or reorganize video content, \eg sorting math topics by ``complexity'' (P13) or visualizing how ``activity level'' changes from passive rest to hands-on preparation when making sourdough (P15).

\subsubsection{Familiarity shaped the form of foraging} 
Although overall feature usage did not vary significantly between familiar
and unfamiliar videos, we observed different qualitative patterns.

With \textbf{familiar videos}, participants used more targeted strategies like semantic search ($M_{f}=1.81$ vs. $M_{u}=1.06$) to refind specific moments.
P11 was impressed how \myquote{I asked nearly every question I had when I first watched the [video] and it gave me everything I wanted. It was easy to find exactly what I was looking for.}
Our visualization also surfaced previously overlooked details ($n=4$, \autoref{fig:likert}: \textit{info finding, discovery, navigation}).
When reforaging a travel video, P1 \myquote{noticed things I didn't catch the first time like the transportation system. It definitely led me to new information.}

For \textbf{unfamiliar videos}, participants expanded more nodes ($M_{f}=12.6$ vs. $M_{u}=14.4$) and explored more unique subtrees ($M_{f}=2.56$ vs. $M_{u}=3.19$), using the hierarchy to quickly assess content and relevance (\autoref{fig:likert}: \textit{overview, understanding}).
Six noted how this overview helped reveal misleading content: \myquote{Seeing this, I feel like I got clickbaited. I was hoping the video would be about modern semiconductors. But it doesn't really talk about it much} (P7).
Similarly, P14 brought a video that seemed lighthearted but was \myquote{actually quite tragic. Wow, I didn't know it was a plane crash. I thought they successfully landed.}

\subsubsection{\edit{Limitations \& future improvements.}}
Users also noted ways to improve \system.
While some found highlights and collections useful for discovering interesting clips, 
four felt these suggestions were too subjective:
\myquote{People have different preferences. [The system] needs more interpretability, like why is that a highlight?}
(P6).

Four saw opportunities for supporting more flexible navigation, for example by \myquote{simultaneously comparing multiple branches} instead of limiting exploration to one subtree at a time (P1).
Relatedly, two proposed decoupling the hierarchy from the video player so \myquote{I [can] watch the video while clicking and finding what to look at next} (P12).
Three participants envisioned extending \system beyond a single video, \eg by visualizing dependencies between videos (P5) or creating collections with segments that span multiple videos (P7).

\subsection{How Do Users Perceive the AI-Generated Video Structure? (RQ3)}\label{sec:rq3}
\edit{Because \system's interactions depend on the AI-extracted hierarchy, we examined whether participants found this structure coherent and useful for navigation.}

\subsubsection{The hierarchy was representative, but could oversegment at deeper levels.}
Most described the extracted video structure as accurate and representative ($n=14$): \myquote{It did a surprisingly good job. I'm impressed with the level of detail} (P10).
P6 added, \myquote{The AI categorized it really well... it matches what I remember [and] you can clearly see the different sections like where is the filler vs. the main vlog.}
However, participants also identified limitations.
P12 felt lower levels of the hierarchy were too granular: \myquote{Everything here is like a one sentence fragment. So I wonder if it's overfitting to the specific sentence [rather than] the overall storyline.}
Three noted minor boundary inaccuracies (P5: \myquote{The segment started in the middle of the question, so I had to go back a bit for context}) or uneven segments (P14: \myquote{I'd probably go for more similar-length clips, but this makes sense semantically}).

\subsubsection{Exposing structure reframed the role of AI in video foraging}
Before using \system, 14 participants primarily \myquote{thought of video AI as just the YouTube question-answering tool. I never considered using AI to break down the video and organize its content like this... so this is really eye-opening} (P1). 
Users also appreciated how \system did not feel like interacting with a chatbot ($n=3$): \myquote{It's using AI to navigate but I don't have to handle that part. It does a lot more than what I could get from prompting ChatGPT myself} (P15).
P16 similarly valued that AI was used to guide rather than replace video viewing: \myquote{I have more of an anti-AI bias for summaries, but here, I like that it's just pointing me to [parts] of a video. All it's doing is saving me time.}

\subsubsection{Users envisioned broader uses of AI-extracted structure}
Beyond foraging, two participants suggested that our AI-generated hierarchy could be applied to language learning: \myquote{My dad doesn't speak much English [but] the AI could break down videos into smaller, edible pieces that he can actually understand and read at his own pace} (P9).
Three saw potential value for content analysis and creation.
P7 said, \myquote{Seeing this structure would be cool if you're trying to understand the choices made in a film. Like how color symbolism moves... or how different forms of media are connected to shape a video essay.}
P12 added, \myquote{I wonder if a creator used this tool, would it cause them to format their videos differently? It's a totally different mode of visualizing things and you're not subject to the linear way the content is initially presented.}

\section{Discussion \& Future Work}
Our findings highlight the value of making AI-extracted video structure visible and interactive to support information foraging.
We reflect on these insights, current limitations, and opportunities for future work.

\subsection{\edit{Supporting Goal-Directed Navigation Through Hierarchical Visualization}}
\edit{With \system, we explored how hierarchical visualization can augment timeline-based navigation for goal-directed viewing.}
While bounded by study scope, our results suggest two primary benefits.
First, surfacing video structure enabled \textbf{faster and more accurate} foraging than \edit{existing workflows} --- and even AI chat (\autoref{sec:rq1}).  
This aligns with Nielsen's ``recognition over recall'' heuristic \cite{budiu2024memory}:
by externalizing structure as a cognitive aid~\cite{scaife1996external,spence1985memory,yeh2026story}, \system allows users to visually identify relevant segments, which is easier than formulating search queries from scratch.
Second, the hierarchy supported foraging in both new and familiar videos by enabling navigation at \textbf{multiple levels of abstraction}.
Rather than relying on repeated linear search, users could skim segments at different granularities based on their information needs~\cite{truong2021automatic,suh2023sense}, which some even described as ``fun'' (\autoref{sec:rq2}).
\edit{Future work should further examine when this flexibility is valuable for information seeking, and when simpler representations are sufficient.}

Beyond navigation, \system's hierarchies helped viewers quickly identify clickbait and misleading content, suggesting that richer structural previews can enable more informed, intentional consumption~\cite{mccrosky2021youtube,chaudhary2022are}. 
Hierarchical summaries have also been shown to improve accessibility~\cite{van2024making}, consistent with participant feedback that surfacing video structure could support uses such as language learning (\autoref{sec:rq3}).
More broadly, this structured visualization approach could help make other complex temporal data more \textit{semantically} navigable, aiding foraging and reforaging across music~\cite{huh2026vidtune,wang2021soloist}, podcasts~\cite{park2024enhancing,wang2024podreels}, and event sequences~\cite{chen2018sequence}. 

\subsection{\edit{Designing Interactions Around AI-Extracted Structure}}
Our work demonstrates how AI-extracted structure, although imperfect (\autoref{sec:rq3}), can enable new interactions with unstructured video beyond conversation.
\edit{To guide future systems built around such structure (in video and beyond), we distill design principles for making these representations useful, appropriately granular, and inspectable.}

\boldni{Ground interactions in interpretable structure.}
Users valued how \system visualized search results in the hierarchy, enabling easy verification compared to isolated text and timestamps (\autoref{sec:rq2}).
In contrast, AI-suggested highlights were harder to interpret, underscoring the need for inspectable outputs linked to source data~\cite{yeh2026story,amershi2019guidelines,wang2024ragviz}.

\boldni{\edit{Adapt structure to (changing) context and intent.}}
\edit{\system currently treats modality as static (\autoref{sec:pipeline}), although the relative importance of visual and textual cues may shift within a video~\cite{wu2020multimodal,yu2022mm}.
Useful structure and scent may also depend on user intent and familiarity: foraging new content may benefit from stronger overview cues, while reforaging may favor deeper access to remembered moments.
Future systems could therefore adapt segmentation depth and surfaced scent as video content and user goals evolve~\cite{wu2022intentvizor,narasimhan2021clip,pu2011user}.}

\boldni{Leverage structure for parallel exploration.}
Selective expansion helped participants focus attention, but they noted how parallel exploration across branches in the hierarchy could support richer video navigation and sensemaking (\autoref{sec:rq2}).
For instance, systems could surface related moments during playback, or expand search and collection results side by side to facilitate comparison and discovery~\cite{suh2023sense,zheng2026t2vtree,liu2025perspectra}.

\subsection{From One to Many: \edit{Foraging and Reforaging} Across Video Collections}
A natural next step is moving beyond single videos to support foraging and reforaging across \textit{collections} of videos~\cite{dal2012multi,hezel2022efficient}, \eg when comparing YouTube results~\cite{perez2022watch,houten2009searching} or learning from multiple lectures~\cite{pavel2014video}.
As the volume of online video --- including increasingly on-demand, AI-generated content --- continues to grow, \edit{users may need new ways to navigate information both within and across videos~\cite{peronikolis2024personalized,van2024making,yeh2026vidmento,zheng2026t2vtree}.}

\edit{Multimodal models now make large-scale video analysis increasingly feasible~\cite{wu2020multimodal,yang2025video}, creating opportunities for AI-augmented foraging interfaces.
While recent work explores semantic retrieval over media collections~\cite{gesuelli2026semantic,huynh2025video}, a key challenge remains in \textbf{visualizing and surfacing connections between videos at scale}. 
From an IFT perspective, each video can be viewed as an information patch with inspectable structure and scent~\cite{pirolli1999information,houten2009searching}.
Systems could use AI to extract and organize these signals across videos, extending \system's semantic dimensions and hierarchical views to compare, filter, and navigate segments at multiple granularities (\autoref{sec:rq2}).
For reforaging, visualizing structural context and adaptive scent could help users recognize and refind previously encountered moments.
Together, these AI-generated structures could transform video collections into interactive information spaces that support goal-directed exploration for both new and familiar content.}

\section{Conclusion}
We introduced \system, an AI-powered browser extension that reframes video navigation as interactive foraging over hierarchical structure.
The system visualizes videos as multi-level semantic roadmaps, surfacing multimodal, adaptive information scent to guide exploration. 
In a user study ($N=16$), \system enabled more efficient and accurate information seeking compared to \edit{existing workflows}.
Participants enjoyed navigating the hierarchical structure at different granularities, 
using it to meet their information needs in both familiar and unfamiliar videos.
Our findings highlight the value of exposing AI-extracted structure for goal-directed viewing.
Beyond individual videos, \system suggests new opportunities for structured visualizations that support foraging and reforaging across video collections and richer sensemaking of other complex temporal data.

\bibliographystyle{ACM-Reference-Format}
\bibliography{references}

\newpage
\appendix
\onecolumn
\renewcommand{\thefigure}{\thesection\arabic{figure}}
\renewcommand{\thetable}{\thesection\arabic{table}}

\section{Prototype Used for Formative Study}\label{app:form_study}
\setcounter{figure}{0}

\autoref{fig:exp_old} and~\autoref{fig:full_old} show the state of our interface at the time of our formative study (\autoref{sec:formative}). 
Both screenshots show a visualization of the following YouTube video: \href{https://www.youtube.com/watch?v=9wRWrSTorLI}{2 Days in Boston: Things to do for First-Time Visitors} by Justin Bernstein (Creative Commons license).

\figureExpOld

\figureFullOld

After these initial interviews, we improved segment styling based on participant feedback (\autoref{fig:exp_old}), and added several core features (\eg adaptive scent, semantic search, collections, contextual color scales).
Participants also noted that the horizontal overlap between expanded layers made it difficult to keep track of the video's chronology, so we adjusted the layout of our final interface to preserve the sequential order of segments.

Another key interface change was removing the ``full view'' of the hierarchical visualization (\autoref{fig:full_old}).
Although users thought it was neat to see the entire video roadmap, they found the visualization overwhelming and challenging to navigate for more complex videos, so we opted to focus on the expandable, progressive disclosure view.
However, based on our study results (\autoref{sec:results}), it may still be valuable to explore alternative ways to surface a more comprehensive video structure to users, or some middle ground between an ``expandable'' and ``full'' hierarchical visualization.

\section{Additional Interface Figures}\label{app:interface}
\setcounter{figure}{0}
We include more figures of \system to illustrate additional system features, including:

\begin{itemize}
    \item \textbf{Semantic Zoom} in \autoref{fig:zoom} and \textbf{Highlights} in \autoref{fig:highlights}, which show a visualization of \href{https://www.youtube.com/watch?v=W3RLUa4nJas}{How to Finally Stop Struggling with Productivity} by \textit{She Makes It Work}.
    \item \textbf{Collections} in \autoref{fig:collections}, which shows a visualization of \href{https://www.youtube.com/watch?v=57_drtPS9h0}{A weekend in Leiden | Netherlands travel vlog} by \textit{Janica's Journey}.
\end{itemize}
Both videos have Creative Commons licenses.

\figureZoom

\figureHighlights

\clearpage
\figureCollections

\section{Extended Comparison Task Results}\label{app:comp_results}
\setcounter{table}{0}

Here, we show the full statistical results from analyzing participant Likert-scale ratings, question accuracies, and completion times after the comparison task (\autoref{sec:results}).

\subsection{Participant Ratings}

There was a significant main effect of interface across all seven Likert-scale questions
($\beta$ ranging from $0.88$ to $1.50$, all $p_{\text{BH}} < .001$), with \system rated
consistently higher than YouTube. 
No interface\,$\times$\,video\_type interaction was
significant (all $p_{\text{BH}} \geq .45$), indicating the advantage of \system was
consistent across both video types.

\tableCompareResults

For visual-driven videos, participants rated \system as easier to find information with
($M_{\text{YT}} = 2.63$ vs.\ $M_{\text{VT}} = 4.38$; $t(12.5) = -3.82$, $p_{\text{BH}} < .01$),
and more helpful for navigating
($M_{\text{YT}} = 3.50$ vs.\ $M_{\text{VT}} = 4.63$; $t(11.0) = -3.00$, $p_{\text{BH}} < .05$)
and understanding videos
($M_{\text{YT}} = 2.88$ vs.\ $M_{\text{VT}} = 4.50$; $t(11.9) = -4.64$, $p_{\text{BH}} < .01$). 
For dialogue-driven videos, participants felt it was easier to obtain an overview of the video
($M_{\text{YT}} = 3.38$ vs.\ $M_{\text{VT}} = 4.63$; $t(14.0) = -3.36$, $p_{\text{BH}} < .05$)
and were more confident in finding information
($M_{\text{YT}} = 2.63$ vs.\ $M_{\text{VT}} = 4.50$; $t(10.3) = -4.47$, $p_{\text{BH}} < .01$)
with \system.

Across all videos, participants saw \system as easier to discover relevant segments with
(dialogue: $M_{\text{YT}} = 2.88$ vs.\ $M_{\text{VT}} = 4.00$; $t(13.1) = -2.55$, $p_{\text{BH}} < .05$; visual: $M_{\text{YT}} = 3.00$ vs.\ $M_{\text{VT}} = 4.63$; $t(10.1) = -3.87$, $p_{\text{BH}} < .01$) and more enjoyable to use (dialogue: $M_{\text{YT}} = 3.50$ vs.\ $M_{\text{VT}} = 4.38$; $t(12.4) = -2.70$, $p_{\text{BH}} < .05$; visual: $M_{\text{YT}} = 3.25$ vs.\ $M_{\text{VT}} = 4.63$; $t(12.8) = -4.44$, $p_{\text{BH}} < .01$).

\subsection{Participant Performance}
We examined participant performance through accuracy and time to answer the four information-seeking questions.
For question accuracies, there was a significant main effect of interface for the synthesis ($\beta=11.46$, $p=.01$) and overall questions ($\beta=11.61$, $p<.001$), meaning participants performed better on the synthesis question, and overall, with \system compared to YouTube.
For completion times, there was a significant main effect of interface for the evidence ($\beta=-21.25$, $p=.045$) and overall completion time ($\beta=-49.75$, $p=.035$), meaning participants were able to answer the evidence question, and finish all the questions, faster with \system.
The interface effect did not significantly vary by video type (all interaction $p \geq.858$ for accuracy, $p \geq.98$ for completion time).

\tableScoreResults

\section{Technical Evaluation}\label{app:tech_eval}
\setcounter{figure}{0}
\setcounter{table}{0}

We conducted an online annotation study to evaluate the quality of the multimodal, hierarchical video summaries generated by our data extraction pipeline (\autoref{sec:pipeline}).

\subsection{Evaluation Setup}

For more details about the list of videos and questions used, along with the survey interface, see the Supplementary Materials.

\subsubsection{Baseline}
In lieu of a comparable \textit{hierarchical} segmentation method with titles, keyframes, and descriptions, we constructed a baseline by repeatedly applying ARC-Chapter \cite{pu2025arc}, an MLLM-based video chaptering model.
As ARC-Chapter does not produce keyframes, one author chose them based on model output without viewing our pipeline's results. 
We adjusted summaries for a fair comparison (\eg limiting text to 1 sentence, and the top layer to 7 segments).

\subsubsection{Materials}
Stimuli included four videos:
two visual-driven (travel vlog, recipe) and two dialogue-driven (product review, advice).
Each video was $\sim$8.5 minutes to scope annotation; we also collected user feedback on longer videos in~\autoref{sec:study}.
For each video, we generated 5 segmentations per method (10 total): one at layer 1, two at layer 2, and two at layer 3.
Layer 2 and 3 summaries were constructed by randomly selecting and splitting the same parent segment for both methods (\eg two layer 1 segments from our pipeline were used for layer 2).

\subsubsection{Method}
1160 judgments were collected from 20 annotators (5 per video) via Qualtrics.
Segmentations were presented in increasing granularity (layer 1 $\rightarrow$ 3), counterbalanced within each layer.
The two summaries for each clip were rated separately (order randomized) on ten criteria via 7-point Likert scales, \eg visual and dialogue coverage, faithfulness, coherence, diversity (adapted from~\cite{gupta2023comprehensive}).

\subsection{Results}
We analyzed ratings using Wilcoxon signed-rank tests.
Mean differences across \textit{video type} and \textit{hierarchy layer} are summarized in~\autoref{fig:eval_ratings}, and the full statistical results are provided in~\autoref{tab:vt_vs_baseline_videotype} (video type analysis) and~\autoref{tab:vt_vs_baseline_layer} (hierarchy layer analysis).

\figurePipelineEval

\tableVideoType

Overall, our pipeline achieved comparable or higher scores than the baseline (overall: $p<.001$).
We saw significant gains in visual (overall: $p<.05$)
and dialogue coverage (visual: $p<.05$, overall: $p<.01$), faithfulness (visual + overall: $p<.05$), coherence (overall: $p<.05$), segment diversity (dialogue: $p<.01$, overall: $p<.05$), correct granularity (dialogue: $p<.05$), and text utility (dialogue: $p<.05$, overall: $p<.01$).
Improvements in image utility were most limited, suggesting that keyframe selection remains a challenging task.
Across the hierarchy, there were no significant differences for layers 1 and 2, but at layer 3, our pipeline significantly outperformed the baseline (all criteria: $p<.05$), illustrating its support for fine-grained segmentation.

\tableLayer

\end{document}